\documentstyle[pre,aps,epsf,multicol]{revtex}
\draft
\begin{document}
\newcommand{\bm}{\bibitem}
\newcommand{\bgea}{\begin{eqnarray}}
\newcommand{\ndea}{\end{eqnarray}} 
\newcommand{\bge}{\begin{equation}} 
\newcommand{\nde}{\end{equation}}
\newcommand{\lbl}{\label}
\newcommand{\rf}[1]{(\ref{#1})}
\newcommand{\srf}[1]{\ref{#1}}

\newcommand{\cc}{\mathop{\rm c.c.}\nolimits}
\newcommand{\hf}{\frac{1}{2}}
\newcommand{\bb}[1]{{\bf #1}}
\newcommand{\rr}{\bb r}
\newcommand{\k}{\bb k}
\newcommand{\vfi}{\varphi}
\newcommand{\vna}{\bb \nabla}
\newcommand{\lpl}{{\bb \nabla}^2}
\newcommand{\ibid}{{\it ibid }}
\newcommand{\prtt}{\partial_t}
\newcommand{\prtr}{\partial_r}
\newcommand{\prtf}{\partial_\phi}
\newcommand{\prtpl}{\partial_{\|}}
\newcommand{\prtpp}{\partial_\perp}
\newcommand{\prtR}{\partial_R}
\newcommand{\prtT}{\partial_T}
\newcommand{\prtF}{\partial_\Phi}
\newcommand{\rb}{Rayleigh-B\'{e}nard }
\newcommand{\ob}{Oberbeck-Boussinesq }
\newcommand{\omrm}{\omega_m R_m}
\newcommand{\q}{\bb q}
\newcommand{\hpsi}{\hat{\psi}}
\newcommand{\hrho}{\hat{\rho}}
\newcommand{\hphi}{\hat{\phi}}
\newcommand{\hchi}{\hat{\chi}}
\newcommand{\heta}{\hat{\eta}}
\newcommand{\hj}{\hat{J}}
\newcommand{\bs}{\bar{S}}
\newcommand{\bg}{\bar{g}}
\newcommand{\sprm}{\hskip -5pt \raisebox{2pt}{${}'$}} 
\newcommand{\sprmm}{\hskip -8pt \raisebox{2pt}{${}'$}}
\newcommand{\sprmmm}{\hskip -11pt \raisebox{2pt}{${}'$}}
\newcommand{\pv}[1]{\langle #1\rangle}
\newcommand{\tv}[1]{\overline {#1} } 
\newcommand{\gpt}{g_{PT}}
\newcommand{\xis}{\xi_{S}}
\newcommand{\xit}{\xi_2}
\newcommand{\km}{k_{max}}
\newcommand{\calf}{{\cal F}}
\newcommand{\te}{{\tilde{\epsilon}}}
\newcommand{\et}{\epsilon_T}
\newcommand{\stc}{spatiotemporally chaotic }
\newcommand{\jsdc}{J_{SDC}}
\newcommand{\osdc}{\Omega_{SDC}}

\newlength{\figsize}
\setlength{\figsize}{0.225\textwidth} 
\newlength{\plotsize}
\setlength{\plotsize}{0.46\textwidth} 
\title{Nature of Roll to Spiral-Defect-Chaos Transition}
\author{Xiao-jun Li$^1$, Hao-wen Xi$^2$ and J. D. Gunton$^1$}
\address{$^1$Department of Physics,
	Lehigh University, Bethlehem, Pennsylvania 18015}
\address{$^2$Department of Physics and Astronomy,
	Bowling Green State University,
	Bowling Green, Ohio 43403}
\date{\today}
\maketitle	
\begin{abstract}
	We investigate the nature of the parallel-roll to 
spiral-defect-chaos (SDC) transition in Rayleigh-B\'enard convection, 
based on the generalized Swift-Hohenberg model. We
carry out extensive, systematic numerical studies 
by, on one branch, 
increasing the control parameter gradually from 
the parallel roll regime to the SDC regime
and, on the other branch, decreasing it in the opposite manner. 
We find that 
the data of several time-averaged {\it global} quantities 
all form hysteretic loops from the two branches.  
We also discuss several possible scenarios for the transition and 
analyze our data for SDC accordingly. We conclude
that the roll-to-SDC
transition is first-order in character and that 
the correlation length diverges at the conduction to convection 
onset. We further postulate 
that this transition can be understood somewhat
similar to the hexagon-to-roll transition in non-Boussinesq fluids. 
Finally we comment on the differences between our conclusion 
and those in two experiments.

\end{abstract}


\begin{multicols}{2}

\section{Introduction} \lbl{int}

Many nonequilibrium systems exhibit self-organized pattern-forming phenomena
\cite{cr_ho_93}.  In the past few years,
a new type of intrinsic pattern has emerged in many disciplines of science.
These patterns are characterized by their extensive, irregular behavior in both
space and time, which are known as 
spatiotemporal chaos (STC) \cite{cr_ho_93}. 
STC typically exists in large size systems and
its complexity increases dramatically with the system size
\cite{gr_96}. Owing to its generic dynamical complexity, STC
hence poses a great challenge to both experimentalists and theoreticians.

	From the 
very beginning, Rayleigh-B\'enard convection (RBC) has been a
paradigm in the study of 
pattern formation in driven dissipative systems, 
because of its relative simplicity and 
high precision in controlled experiments \cite{ah_95}.
RBC can occur when 
a thin horizontal layer of fluid is heated from below. The system 
is described by
three dimensionless parameters \cite{cr_ho_93}: (a)  
the Rayleigh number $R \equiv g \alpha d^3 \Delta T/\kappa \nu$,
in which $g$ is the gravitational acceleration,
$d$ the layer thickness, $\Delta T$ the temperature gradient 
across the layer,  
$\alpha$ the thermal expansion coefficient, $\kappa$ the thermal diffusivity 
and $\nu$ the kinematic viscosity; (b)    
the Prandtl number $\sigma \equiv \nu/\kappa$; and, (c) 
the aspect ratio 
$\Gamma \equiv L/2d$ where $L$ is the horizontal size of the system.
The Rayleigh number $R$ is the control parameter of the system; the
Prandtl number $\sigma$ specifies the fluid properties. 
It is convenient to introduce a reduced control parameter 
$\epsilon \equiv (R - R_c)/R_c$, where $R_c$ is the critical value of $R$
at which the fluid bifurcates from a static conductive state to a convective
state.  

RBC has been studied extensively in the literature 
\cite{cr_ho_93,ah_95}. Theoretical
analyses by Busse and his coauthors \cite{sc_lo_65} predict that parallel roll
states are stable inside a stability domain in $(R, k, \sigma)$ space with
$k$ the wavenumber, which is known as the ``Busse balloon''. 
Surprisingly, 
recent experiments \cite{mo_bo_93,as_st_93} and numerical studies 
\cite{xi_gu_93,be_fa_93,de_pe_94}, using systems with $\sigma \sim O(1)$
and large $\Gamma$, revealed that 
the parallel roll state yields to a 
spatiotemporally chaotic  state even for states inside the 
Busse balloon. 
This \stc state, called {\it spiral-defect-chaos} (SDC),   
exhibits very complicated dynamics, illustrated by the interplay of 
numerous rotating spirals, patches of moving rolls, 
intricate grain boundaries, 
dislocations and other defects \cite{mo_bo_93,as_st_93}. 
Its discovery 
has since stimulated many experimental 
\cite{mo_bo_93,as_st_93,hu_ec_95,mo_bo_96,ca_eg_97}, 
theoretical \cite{cr_tu_95,li_xi_97} 
and numerical \cite{xi_gu_93,be_fa_93,de_pe_94,xi_gu_95} 
efforts to understand it. 

The nature of the parallel roll to SDC transition is one of the 
important questions with respect to SDC and 
has been investigated in several experimental and numerical studies
\cite{mo_bo_96,ca_eg_97,xi_gu_95}. 
By solving the  generalized Swift-Hohenberg (GSH) model of RBC 
\cite{sw_ho_77,cr_80,si_zi_81} for non-Boussinesq fluids with {\it random} 
initial conditions, two of us  \cite{xi_gu_95}
characterized the transition by the 
behavior of time-averaged
global quantities such as convective current $J$, vorticity
current $\Omega$ (called vortex energy in Ref. \cite{xi_gu_95}) 
and spectra entropy $\Xi$ \cite{po_pe_79}.
It was found that the convective current seems to be smooth across the 
transition temperature $\et$
but both the vorticity current and the spectra entropy seem to obey
power-law behavior near $\et$. However this study was unable to distinguish
between a gradual or sharp transition.  
Despite its inconclusiveness on the nature of the roll-to-SDC 
transition,  this study suggests that 
studying such time-averaged global quantities is quite useful. 
It hence motivated us to 
develop a phenomenological theory for STC, 
including SDC, in RBC \cite{li_xi_97}. In the theory,
we made a random phase approximation for the spatiotemporally chaotic states
and assumed that their time-averaged
structure factor $S(k)$ satisfies a scaling 
form with respect to the two-point correlation length $\xit$. With these
assumptions, 
we obtained analytical expressions for both $J$ and $\Omega$
in terms of measurable quantities. These theoretical results provide us  
some new insights on the nature of SDC. In addition,  
a recent experimental study \cite{ca_eg_97} also found 
spectra entropy $\Xi$ to be a useful quantity. 

On the experimental side, Morris {\it et al.} \cite{mo_bo_96}
studied the structure of
SDC using a {\it circular} cell. They found that the correlation length $\xit$
is smooth across $\et$ and diverges at $\epsilon = 0$ with a mean-field
exponent. But the data for the correlation time $\tau$ was consistent with
either a divergence at $\et$
with a mean-field exponent or a divergence at $\epsilon = 0$ with a 
non-mean-field exponent. 
However, a different conclusion was reached 
by Cakmur {\it et al.}
\cite{ca_eg_97} recently, who used a {\it square} cell in their experiment. 
They found that $\xit$ diverges at $\et$
with a small exponent. 
In critical phenomena, we know that 
finite (infinite) $\xit$ and $\tau$ are normally associated with  
first-order (second-order) transitions.  
If this is also true in non-equilibrium phenomena, then these two experiments
should lead to 
different conclusions  about the nature of the roll-to-SDC transition at $\et$.
Although it is possible that the nature of the transition is different
for systems with different geometry, as suggested in Ref. \cite{ca_eg_97},
it is not clear whether this is true in the limit of 
an infinite system. 
We believe, however, that  
the results of Cakmur {\it et al.} \cite{ca_eg_97}
should be treated with caution, particularly given the absence of 
data for $\xit$ for the parallel roll states (i.e., $\epsilon < \et$) and the
relatively limited number of data near $\et$.  
We will comment further about these two experiments in Sec. \srf{conc}. 

In this paper, 
we present our extensive, systematic numerical studies of SDC, based on the 
GSH model of RBC for Boussinesq fluids. In comparison with studies  
in Ref. \cite{xi_gu_95}, we use random initial conditions 
at $\epsilon = 0.05$ and $\epsilon = 0.8$ only. Then, after completing
the calculation at one $\epsilon$,  
we increase $\epsilon$ 
(originally 
from $\epsilon = 0.05$) or decrease $\epsilon$ (originally 
from $\epsilon = 0.8$)
gradually and  
take the final state from the previous
$\epsilon$ as our initial condition. 
We hence obtain two different branches of data, one from increasing 
$\epsilon$  and the other from decreasing $\epsilon$.
We find that the results for $\xit$, $J$, $\Omega$ and $\Xi$ 
all form hysteretic loops from the two branches during the roll-to-SDC 
transition. We analyze our data in accordance with 
our theoretical results \cite{li_xi_97}. 
We conclude that the roll-to-SDC transition is first-order in character.
We also postulate that this transition can be understood somewhat similar to
the hexagon-to-roll transition in non-Boussinesq fluids \cite{bu_67}; namely,
we postulate that the SDC 
bifurcation actually occurs at $\epsilon =0$. But, since SDC is  
unstable (or metastable)
against parallel roll states at smaller $\epsilon$, it  emerges 
only for $\epsilon > \et$.

This paper is organized as follows.
In Sec. \srf{th}, we introduce the GSH model of RBC and define  
some important time-averaged global quantities.  We then summarize our
theoretical results in Ref. \cite{li_xi_97}. In Sec. \srf{sce}, we  
discuss possible scenarios with regard to SDC, according to whether
the correlation length $\xit$ 
diverges at $\epsilon = 0$ or at $\et$.
We present the details of our numerical studies in Sec. \srf{num}. We also
analyze
the data for $\xit$, $J$, $\Omega$ and $\Xi$ to test these different
scenarios.
In the last section, we discuss the subtleties 
involved in determining the nature
of the roll-to-SDC transition and comment on the different
conclusions between this work and  the earlier
experimental studies \cite{mo_bo_96,ca_eg_97}.

\section{Theoretical Results} \lbl{th}

The GSH model of RBC 
\cite{sw_ho_77,cr_80,si_zi_81} 
is widely accepted for theoretical study. 
This model is derived from the three-dimensional hydrodynamic equations,
but is much simpler to study both numerically and analytically. 
The GSH model contains  
two coupled equations in two-dimensional space $\rr = (x,y)$,   
one for the order parameter  $\psi(\rr,t)$ and the other for the
mean-flow field $\zeta(\rr,t)$.
The convective patterns in RBC are 
completely determined by the order parameter $\psi(\rr,t)$.
The amplitude equations for the GSH model and the hydrodynamical equations
are the same in the leading order near onset. 
Numerical solutions of this model or its modified versions 
have not only reproduced most patterns observed in 
experiments but also resembled experimental results relatively well
\cite{cr_ho_93,xi_gu_93,be_fa_93,cr_tu_95,xi_gu_95,xi_vi_92}. 
But there are some 
shortcomings in the model \cite{de_pe_94,gr_cr_85}: The stability boundary
of the model does not coincide with that of hydrodynamics; it induces
an unphysical, short-ranged cross roll instability; and both the shape and the
peak position of the power spectrum for SDC are different from those in
the real system. Even so, 
owing to its simplicity and its qualitative resemblance to real systems,
this model is very valuable in studying RBC. 

In the GSH model,
the order parameter $\psi(\rr,t)$ satisfies \cite{sw_ho_77,cr_80,si_zi_81}  
\bge
         \partial_{t} \psi + g_m \bb U \cdot \bb \nabla \psi
        = \left[\epsilon - ({\bb \nabla}^2 + 1)^2 \right] \psi - \psi^3, 
		\lbl{gsh}
\nde
where $\nabla$ is the gradient operator in two-dimensions and 
$\bb U(\rr)$ is the mean-flow velocity given by
$\bb U(\rr) = \vna \zeta(\rr,t) \times {\bb e}_z$. The mean-flow field
$\zeta(\rr,t)$, on the other hand, satisfies \cite{si_zi_81}
\bge
        \left[\partial_{t}-\sigma ({\bb \nabla}^2 - c^2)\right] 
		{\bb \nabla}^2 \zeta
        = {\bb e}_z \cdot \left[{\bb \nabla}( {\bb \nabla}^2 \psi)
		\times {\bb \nabla} \psi\right].
		\lbl{mf}
\nde
Variables in these equations have been
rescaled for numerical convenience. Their relations to their 
physical values can be found  in Ref. \cite{li_xi_97}.
[See Eqs. (1) - (3) and (58) - (60) there.] 
For example, the reduced Rayleigh number $\epsilon$ in Eq. \rf{gsh}
is related to its physical value $\epsilon_{\rm expt}$ by 
$\epsilon_{\rm expt} = 0.3594 \epsilon$. The rescaling factors for 
$\psi$, $\zeta$, $t$ and $\rr$ and 
parameters $g_m$, $\sigma$ and $c^2$ can also 
be found in Ref. \cite{li_xi_97}. 

We now define several important time-averaged {\it global} quantities in RBC. 
The first one is 
the time-averaged convective current defined as  
\bge
	J \equiv A^{-1} \int d \rr\,\tv{\psi^2(\rr,t)} 
	= \sum_{\k} \tv{\hpsi^*(\k,t) \hpsi(\k,t)}, \lbl{cvcrnt}
\nde 
where $\hpsi(\k,t)$ is the Fourier component of $\psi(\rr,t)$, 
$A$ is the area of the system
and $\tv{F(t)}$ represents the time-average of $F(t)$. This quantity 
increases from $J=0$ to $J > 0$ at the conduction-to-convection onset and hence
characterizes the transition. 
The second one is 
the time-averaged vorticity current defined as 
\bge 
\Omega \equiv A^{-1} \int d \rr\,\tv{\omega_z^2(\rr,t)}, \lbl{vtctcrnt}
\nde
where $\omega_z(\rr,t) = - {\bb \nabla}^2 \zeta(\rr,t)$. 
This quantity reflects the distortion of patterns at large distance. It 
is identically zero for perfect parallel rolls and increases dramatically 
for SDC \cite{xi_gu_95}. Because of this, it has been speculated that one may
take $\Omega$ as one of the order parameters to characterize the roll-to-SDC
transition \cite{xi_gu_95}.  The third one is the time-averaged spectra entropy
defined as \cite{po_pe_79} 
\bge
	\Xi  \equiv - \sum_\k \tv{S(\k,t) \ln S(\k,t)}, \lbl{entropy}
\nde
where the structure factor  $S(\k,t)$ is given by
\bge
	S(\k,t) = \hpsi^*(\k,t) \hpsi(\k,t)/J. \lbl{strct_fct}
\nde
Like its counterpart in thermodynamics, the spectra entropy $\Xi$ 
is related to the randomness of all excited states. Its value  
is $\ln 2 = 0.6931$ for perfect 
parallel roll states but increases dramatically for SDC. This quantity
was also found useful to characterize the roll-to-SDC transition 
\cite{ca_eg_97,xi_gu_95}. Another important quantity is 
the two-point correlation length defined as 
\bge
	\xi_2 = \left[\pv{k^2}_k - \pv{k}_k^2\right]^{-1/2}, \lbl{lngth}
\nde
where we has used the notation 
$\pv{F(k)}_k = \int^\infty_0 dk \, kS(k) F(k)$ in which 
$S(k)$ is the time-averaged and azimuthally averaged structure factor 
normalized by $\int^\infty_0 dk \, kS(k) = 1$. Clearly $\xit$ specifies the
width of $S(k)$.  

In Ref. \cite{li_xi_97} we presented our analytical calculations,
using the GSH model,  of $J$ and $\Omega$ 
for STC in RBC. These calculations are valid for both 
SDC and phase turbulence (PT) \cite{bu_89,xi_li_97}. 
By assuming the time-averaged two-point correlation function 
\bge
	C(\rr_1,\rr_2) 
	\equiv \tv{\psi(\rr_1,t) \psi(\rr_2,t)}/\tv{\psi^2(\rr_1,t)}, 
	\lbl{crr_fn}
\nde 
is translation invariant in STC, i.e., $C(\rr_1,\rr_2) = C(\rr_1 - \rr_2)$, 
we found that the phases of two $\hpsi(\k,t)$ fields are 
uncorrelated in time
unless they have the same wavenumber $\k$. Furthermore, we 
applied a {\it random phase approximation} (RPA) to STC in which
four-point correlation functions are approximated by products of
two-point correlation functions such as 
$\tv{\psi \psi \psi \psi} \sim  \tv{\psi \psi}\, \tv{\psi \psi}$.
Using this RPA, we derived 
$J$ and $\Omega$ in  
terms  of $S(k)$.  
We further 
assumed that the structure factor satisfies a {\it scaling} form with respect
to $\xit$, i.e., 
\bge
	k S(k) = \xi_{2} \calf[(k - k_{max}) \xi_{2}],  \lbl{s_scaling}
\nde
where $\km$ is the peak position of $k S(k)$
and  $\calf(x)$ is the scaling function satisfying 
$\int_{-\infty}^\infty d x\, \calf(x) = 1$.  
[Since $k \ge 0$ in $k S(k)$,
the lower limit for $\calf(x)$ is $- k_{max} \xi_2$, which we approximate
by $-\infty$.]
From these assumptions, 
we obtained explicit formulas for both $J$ and $\Omega$ in the leading order
of $\xit^{-1}$. For SDC, these results can be written as \cite{li_xi_97}
\bge
	J^{SH}_{SDC} \approx 
	\frac{2}{3}\left[\epsilon  - \frac{4 \pv{x^2}_x}{\xi^2_2}\right], 
	\lbl{tvcvcrnt}
\nde 
and
\bge
	\Omega^{SH}_{SDC} \approx \frac{1}{2 \sigma^2}
	\left[\frac{2 + c^2}{\sqrt{4 c^2 + c^4}} -1\right] 
	\frac{(J_{SDC}^{SH})^2}{\xit^2}, 
	\lbl{tvvtctcrnt}
\nde
where we have used the notation
$\pv{F(x)}_{x}=\int_{-\infty}^\infty d x \, \calf(x) F(x)$. 
Inserting $k = \km + x \xi_2^{-1}$ 
and Eq. \rf{s_scaling} into Eq. \rf{lngth},
it is easy to see that 
$\pv{x^2}_x = 1 + \pv{x}_x^2 \ge 1$ and  
$\pv{k}_k = \km + \xit^{-1} \pv{x}_x$. 

In comparison, the convective current 
for perfect parallel rolls with wavenumber $k_0$ has been evaluated 
to be 
$J^{SH}_{Roll} = \frac{2}{3} [\epsilon - (1 - k_0)^2]$ \cite{cr_80}. 
To our knowledge, there is no explicit formula for $J^{SH}_{Roll}$ 
for distorted rolls.  
If one uses this expression for $J^{SH}_{Roll}$ 
but replaces $(1-k_0)^2$ with $\pv{(1-k)^2}_k$ to account for the
finite width of the power spectrum,
one finds for distorted roll states that 
\bge
	J^{SH}_{Roll} 
	= \frac{2}{3} \left[\epsilon - \frac{4 \pv{x^2}_x}{\xit^2}
	\right], 
			\lbl{J_2}
\nde
where $\xit$ and $\pv{x^2}_x$ have exactly the same meanings as in 
Eq. \rf{tvcvcrnt},  but their values may be different. 

Notice that  
the formulas for $J^{SH}_{SDC}$ and $J^{SH}_{Roll}$ are the same. 
This, however, is due to the simplification that the 
coupling constant of the nonlinear
term, $\psi^3$, is taken as a constant in Eq. \rf{gsh}. 
In a more realistic description of hydrodynamics, 
this coupling constant, say $g(\cos \alpha)$, is
angle dependent and has been evaluated in Ref. \cite{cr_80} 
[before the rescalings leading to Eq. \rf{gsh}]. 
For such a coupling constant, the time-averaged convective current for SDC,
before the rescalings, has been calculated in Ref. \cite{li_xi_97} as  
\bge
	J_{SDC} \approx \frac{2}{g_{SDC}} \left[\epsilon 
	- \frac{\pv{x^2}_x \xi_0^2}{\xit^2}\right],\lbl{tvcvcrnt_real}
\nde
where $\xi_0^2 \simeq 0.148$ \cite{cr_80} and
$g_{SDC} = 1.1319 + 0.0483 \sigma^{-1} + 0.0710 \sigma^{-2}$ \cite{li_xi_97}.
Correspondingly, the convective current for parallel rolls is
\bge
	J_{Roll} = \frac{1}{g_{Roll}} \left[\epsilon 
	- \frac{\pv{x^2}_x \xi_0^2}{\xit^2}\right],\lbl{J_2_real}
\nde
with $g_{Roll} = 0.6995 - 0.0047 \sigma^{-1} + 0.0083 \sigma^{-2}$ 
\cite{cr_80}. 
From these expressions, one finds that
\bgea
	\Delta J &\equiv& J_{SDC} - J_{Roll} 
	= \left[\frac{2}{g_{SDC}} -\frac{1}{g_{Roll}}\right] \epsilon
	\nonumber \\ 
	&&
	-\frac{2 \xi_0^2}{g_{SDC}}\left[\frac{\pv{x^2}_x}{\xit^2}\right]_{SDC}
	+\frac{\xi_0^2}{g_{Roll}}\left[\frac{\pv{x^2}_x}{\xit^2}\right]_{Roll}.
	\lbl{d_j}
\ndea
Since $2/g_{SDC}$ is not equal to $1/g_{Roll}$ 
for most values of $\sigma$,  the first term above is not zero.
It is highly unlikely that this term 
may be cancelled by the contributions from the two other terms.
Thus, there exist discontinuities in the value and the slope of $J$ 
during the roll-to-SDC transition. 
This is not surprising considering that $J$ depends sensitively
on the structure of the convective pattern \cite{cr_80} and that the
structures of parallel rolls and SDC are so different. 
Assuming $[\pv{x^2}_x/\xit^2]_{SDC} = [\pv{x^2}_x/\xit^2]_{Roll}$ at the
transition temperature,
we find that $\Delta J/J_{Roll} = 0.1239$ for $\sigma = 1$. So
the value and the slope of $J$ jump 
about $10\%$ during the roll-to-SDC transition.
Similar jumps can be found for other values of $\sigma$ under the same
assumption; see Fig. 1.

\narrowtext
\begin{figure}[t]
\epsfxsize = \plotsize
\epsfysize = 0.67 \plotsize
\centering
\hbox{\raisebox{0.94\epsfysize}{$\frac{\Delta J}{J_{Roll}}$} \hskip -0.2in
\epsfbox{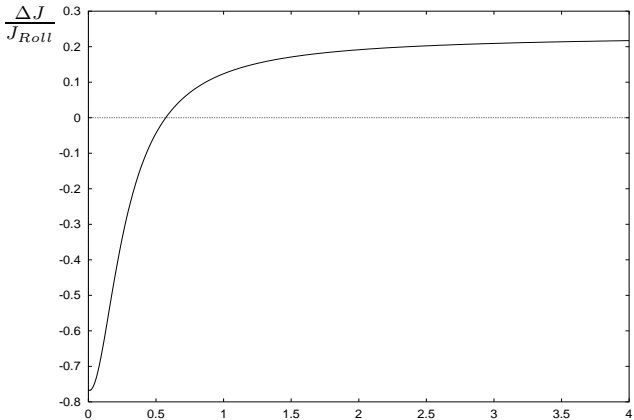}}

\hskip 0.1\plotsize
$\sigma$
\caption{$\Delta J/J_{Roll}$ vs. Prandtl number $\sigma$
where $\Delta J = J_{SDC} - J_{Roll}$.} 
\lbl{fig_jump}
\end{figure}

\section{Possible Scenarios} \lbl{sce}

From Eqs. \rf{tvcvcrnt} and \rf{tvvtctcrnt}, it is obvious that the behavior
of $J_{SDC}$ and $\Omega_{SDC}$ depend sensitively on the two-point 
correlation length $\xit$. For simplicity, we drop the superscripts in 
Eqs. \rf{tvcvcrnt} - \rf{J_2}.
We assume that $\xit$ has a power-law behavior
such as 
\bge
	\xit \approx \xi_{2,0} \te^{-\nu}, \lbl{lngth_asmp}
\nde
where $\te$ is the basic scaling field for SDC. Similar behavior has been
found for $\xit$ in phase turbulence (PT) \cite{li_xi_97}. In that case, since
the transition to PT occurs at $\epsilon = 0$ \cite{bu_89,xi_li_97}, 
one has simply $\te = \epsilon$ \cite{li_xi_97}. 
The possible scenario for SDC, however, is more subtle since the roll-to-SDC
transition occurs at a positive temperature $\epsilon_T$ 
\cite{mo_bo_93,as_st_93,xi_gu_93}. 
One obviously has two alternative choices for
the scaling field $\te$ in Eq. \rf{lngth_asmp}: (A) $\te = \epsilon$ or 
(B) $\te = \epsilon - \epsilon_T$. 
The case (A) is similar to the situation in the hexagon-to-roll transition
\cite{bu_67} 
where the two-point correlation length $\xit$ is finite at the transition 
temperature $\et$ but diverges at $\epsilon = 0$. This scenario is consistent
with the experimental result by Morris {\it et al.} \cite{mo_bo_96}.  
The case (B) resembles the situation in critical phenomena in which
$\xit$ diverges at $\epsilon_T$. 
This scenario was suggested by Cakmur {\it et al.} \cite{ca_eg_97}. 
We now discuss these two scenarios  separately. 

(A) If $\te = \epsilon$. 
This implies 
that all properties of SDC are controlled at $\epsilon = 0$ rather than at 
$\epsilon = \epsilon_T$. 
As far as scaling relations are 
concerned, this case is similar to the situation in PT \cite{li_xi_97}. 
Similar to those in PT, one may hence define power laws such as
\bge
	J_{SDC} \approx J_0 \epsilon^{\mu} \quad {\rm and} \quad 
	\Omega_{SDC} \approx \Omega_0 \epsilon^{\lambda}. \lbl{a1}
\nde
One finds from Eqs. \rf{tvvtctcrnt}
and \rf{lngth_asmp} the following scaling relation 
\bge
        \lambda = 2 \mu + 2 \nu. 
                \lbl{scaling_a1}
\nde
In comparison, one has $\lambda = 2 \mu + \nu$ in PT \cite{li_xi_97}.
From Eq. \rf{tvcvcrnt}, 
since $\jsdc$ is positive by definition, the values of the exponents
satisfy
\bge
	\mu = 1, \quad \nu \ge 1/2 \quad
	{\rm and} \quad \lambda = 2 + 2 \nu \ge 3.
		\lbl{a_expnts}
\nde

Now it is useful to further distinguish two different cases: 
(A1) $\nu = 1/2$, 
or (A2) $\nu > 1/2$. Case (A1) corresponds to a mean-field
exponent $\nu$, in which one has that 
\bge
	J_0 = \frac{2}{3}\left[1 - \frac{4 \pv{x^2}_x}{\xi_{2,0}^2}\right],
	\lbl{a1_j_amp}
\nde
and
\bge
        \Omega_0 =\frac{1}{2 \sigma^2}
	\left[\frac{2 + c^2}{\sqrt{4 c^2 + c^4}} -1\right] 
	\frac{J_0^2}{\xi_{2,0}^2}. 
        \lbl{a1_omega_amp}
\nde
The amplitudes $J_0$ and $\Omega_0$ depend on two
phenomenological parameters $\xi_{2,0}$ and $\pv{x^2}_x$. 
In case (A2),  the exponent $\nu$ has a non-mean-field value.  
Now, since $2 \nu > 1$,  the $\xit^{-2} \sim \epsilon^{2 \nu}$ term
in Eq. \rf{tvcvcrnt} only adds a correction to the leading singularity.  
Instead of Eq. \rf{a1}, one may define
\bge
	\jsdc = J_0 \epsilon^\mu [1 + j_1 \epsilon^{\mu_1} + \cdots],
	\lbl{a2_j}
\nde
and
\bge
	\osdc = \Omega_0 \epsilon^\lambda [1 
		+ \omega_1 \epsilon^{\lambda_1} + \cdots], \lbl{a2_omega}
\nde
with $\mu_1 = 2 \nu - \mu >0$ and $\lambda_1 >0$. 
Consequently, instead of Eq. \rf{a1_j_amp}, one has that 
\bge
	J_0 = 2/3  \quad  {\rm and} \quad j_1 = -4 \pv{x^2}_x/\xi^2_{2,0}.
	\lbl{a2_j_amp}
\nde
While $\Omega_0$ is still given by Eq. \rf{a1_omega_amp}, one must use
the corresponding new value of $J_0$. 
The values of $\omega_1$ and $\lambda_1$, 
however, cannot be determined without knowing the behavior of $\xit$ beyond
the leading term described in Eq. \rf{lngth_asmp}.

(B) If $\te = \epsilon - \epsilon_T$. As we mentioned, 
this case resembles the situation in critical phenomena in which
$\xit$ diverges at $\et$.
Then, as in critical phenomena,
other quantities should also have singular behaviors at $\epsilon_T$.
Now, instead of Eq. \rf{a1} or Eqs. \rf{a2_j} and \rf{a2_omega}, 
we define, near $\te = 0^+$, that  
\bge
	\jsdc \approx J_0 \epsilon - J_{s,0} \te^\mu \quad {\rm and} \quad 
	\osdc \approx \Omega_{s,0} \te^\lambda. \lbl{b}
\nde
The behavior of $\jsdc$ is apparently 
dominated by the smooth background term $J_0 \epsilon$ near $\et$.
As a consequence, the $\jsdc^2$ factor in Eq. \rf{tvvtctcrnt} no longer
contributes to the value of $\lambda$. 
From Eqs. \rf{tvcvcrnt}, \rf{tvvtctcrnt} and \rf{lngth_asmp},
one gets that
\bge
	\mu =  \lambda = 2 \nu, \lbl{b_expnts}
\nde
instead of Eq. \rf{a_expnts}.
As for the amplitudes, one finds that
\bge
	J_0 = \frac{2}{3} \quad {\rm and} \quad 
	J_{s,0} = \frac{8 \pv{x^2}_x}{3 \xi_{2,0}^2}, \lbl{b_j_amp}
\nde
and
\bge
        \Omega_{s,0} = \frac{1}{2 \sigma^2}
	\left[\frac{2 + c^2}{\sqrt{4 c^2 + c^4}} -1\right] 
	\frac{J_0^2 \et^2}{\xi_{2,0}^2}, 
        \lbl{b_omega_amp}
\nde
where we have used $\epsilon \approx \et$ near $\te = 0^+$.
If $2 \nu$ is not an integer, then the best way to evaluate 
the exponents $\mu$ and $\nu$ is, in principle, to
differentiate $\jsdc$ and $\osdc$ with respect to $\epsilon$ and to analyze
the corresponding divergences
after certain orders of differentiation. 
The scaling relations $\mu = \lambda = 2 \nu$, hence, 
provide  a very strong test for 
the $\te = \epsilon - \et$ assumption.
If $\mu = 2 \nu < 1$, then the slope of $J_{SDC}$ is negative in the range of
$0 < \te < \te_\times$ with $\te_\times = [\mu J_{s,0}/J_0]^{1/(1-\mu)}$.
The observation of a negative slope of $J_{SDC}$ near $\et$ will apparently
increase the validity of scenario (B). But if the value of
$\te_\times$ is very small, such an observation may not be practical at present.

\section{Numerical Solutions and Data Analyses} \lbl{num}

We now present our numerical studies of SDC with the GSH equations.
The numerical method for solving the GSH equations
is based on the work by Bj{\o}rstad {\it et al.} \cite{gr_co_84}.
Following Ref. \cite{xi_gu_93}, 
we  choose $g_m=50$, $\sigma=1.0$ and $c^{2}=2.0$ for parameters in 
Eqs. \rf{gsh} and \rf{mf}.
In our simulation, we  take a square cell of size
$L_x=L_y=128 \pi$, which corresponds to an aspect ratio
$\Gamma = 64$. Uniform square grids with
spacing  $\Delta x = \Delta y =\pi /4.0$ have been used, 
so  the total number of nodes is $512 \times 512$.
We use the rigid boundary
conditions $\psi|_{B} = \bb n \cdot \bb \nabla \psi |_{B} =
\zeta|_{B} = \bb n \cdot \bb \nabla \zeta |_{B} = 0$
in the simulation. Here $\bb n$ is
the unit vector normal to the boundary, say $B$, of the domain of integration.
We take two different routes to systematically 
study the transitions between parallel roll states and SDC states: (A)  
We increase the control parameter $\epsilon$ from $\epsilon = 0.05$ to
$\epsilon = 0.6$ with steps of $\Delta \epsilon = 0.05$. We call this the
roll branch. (B) We decrease
$\epsilon$ from $\epsilon = 0.8$ to $\epsilon = 0.4$.
We call this the SDC branch. 
For $\epsilon = 0.05$ or $0.8$, 
we choose initial conditions 
$\zeta(\rr,t=0)$ and $\psi (\rr,t=0)$ as random
variables, obeying a Gaussian distribution with a zero mean and 
a variance of $0.001$. For other subsequent $\epsilon$'s, we take the final
results 
from the previous $\epsilon$ as our initial conditions. For each $\epsilon$, 
we wait about four horizontal diffusion time $t_h$  
before collecting data which, we hope, is sufficient to pass the transient
regime. 
We run for  an additional $20.8 t_h$ to collect $20$ instantaneous profiles 
for SDC states 
or $16 t_h$ to collect $10$ profiles
for parallel roll states during each data collection.  

\begin{figure}[t]
\centering
\begin{tabular}{cc}
\epsfxsize = \figsize \epsfysize = \epsfxsize 
\epsfbox{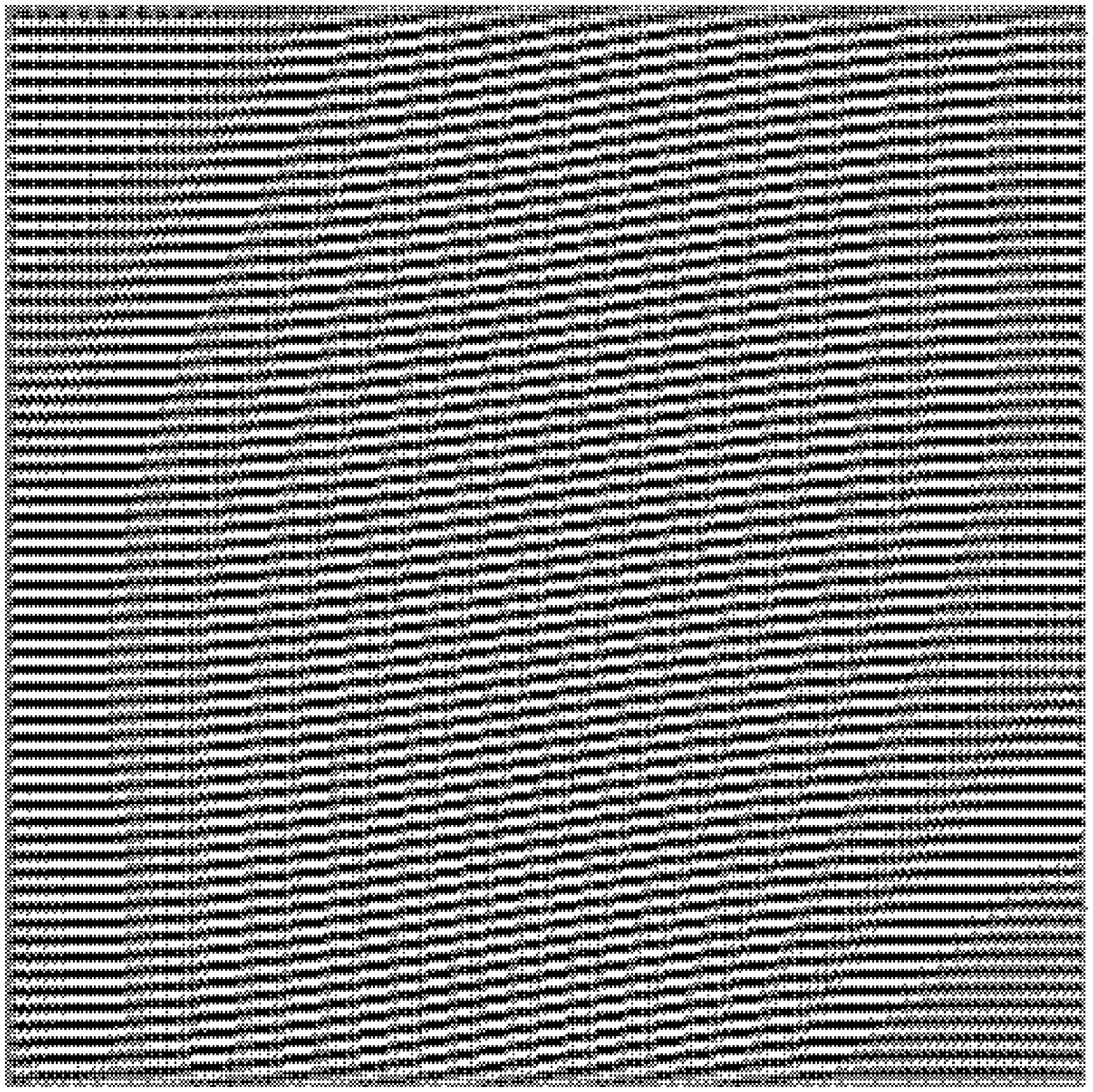} &
	\epsfxsize = \figsize \epsfysize = \epsfxsize 
	\epsfbox{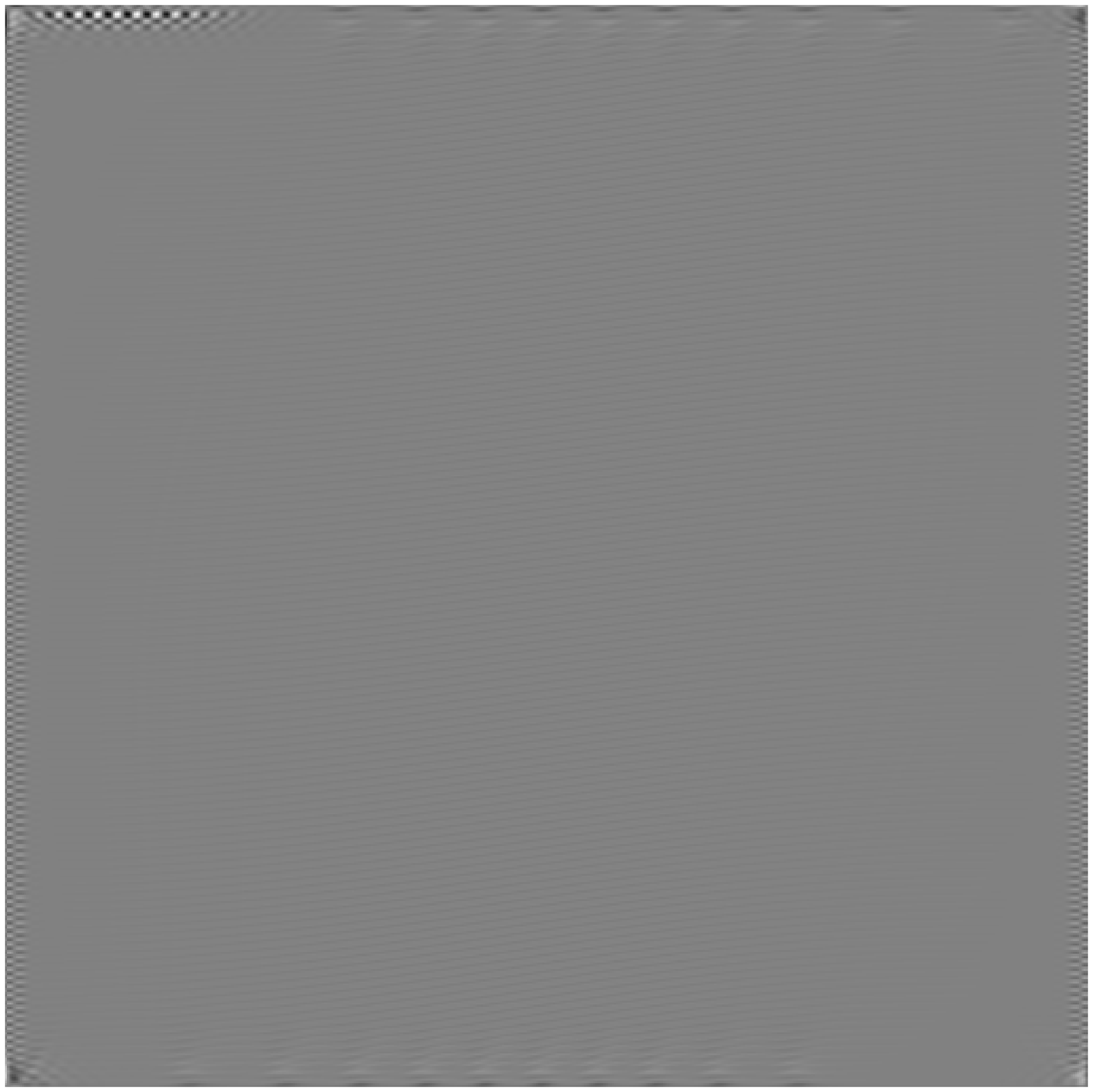} \\
(a) & (b) \\ 
\epsfxsize = \figsize \epsfysize = \epsfxsize \epsfbox{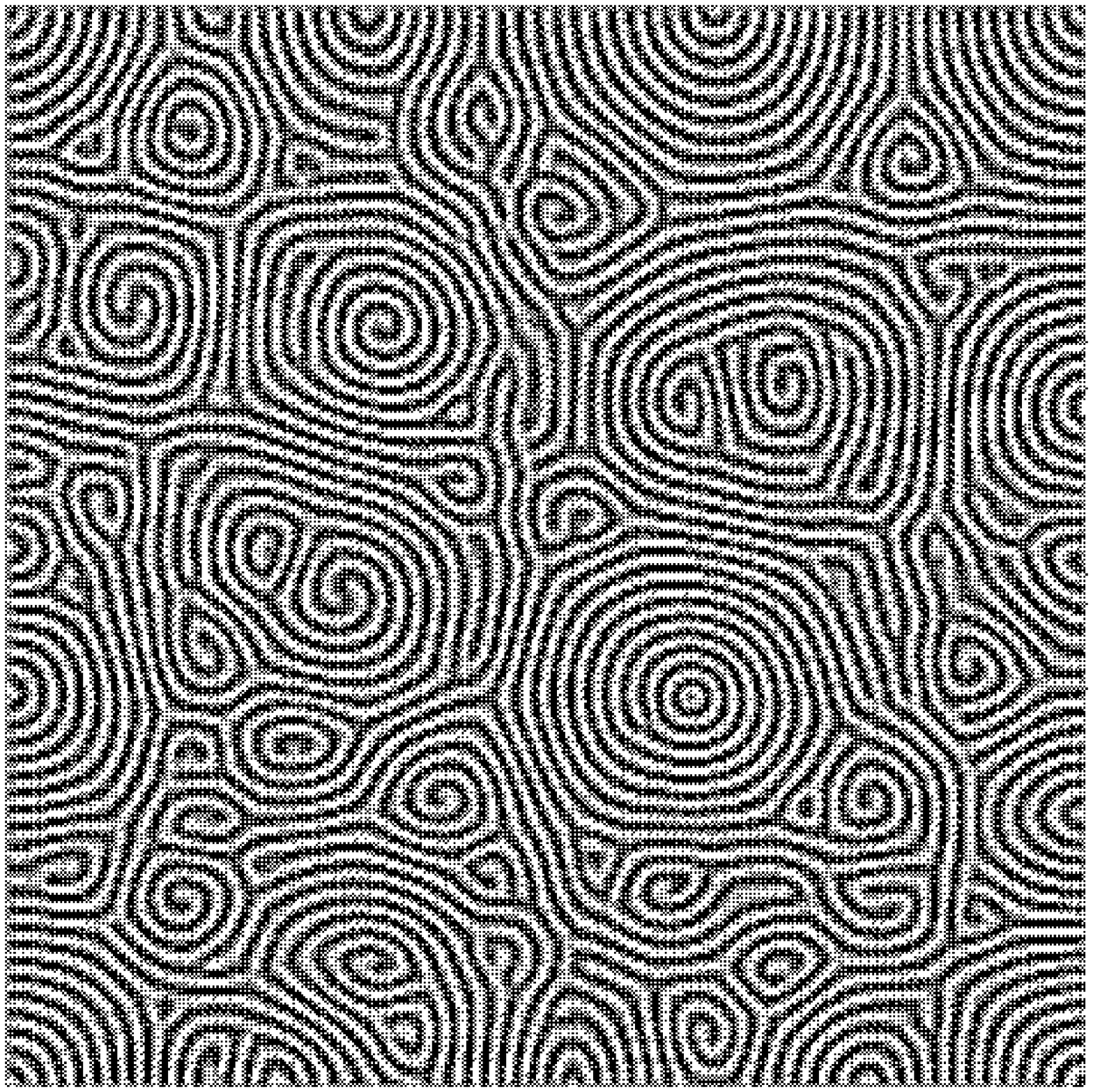} &
	\epsfxsize = \figsize \epsfysize = \epsfxsize 
	\epsfbox{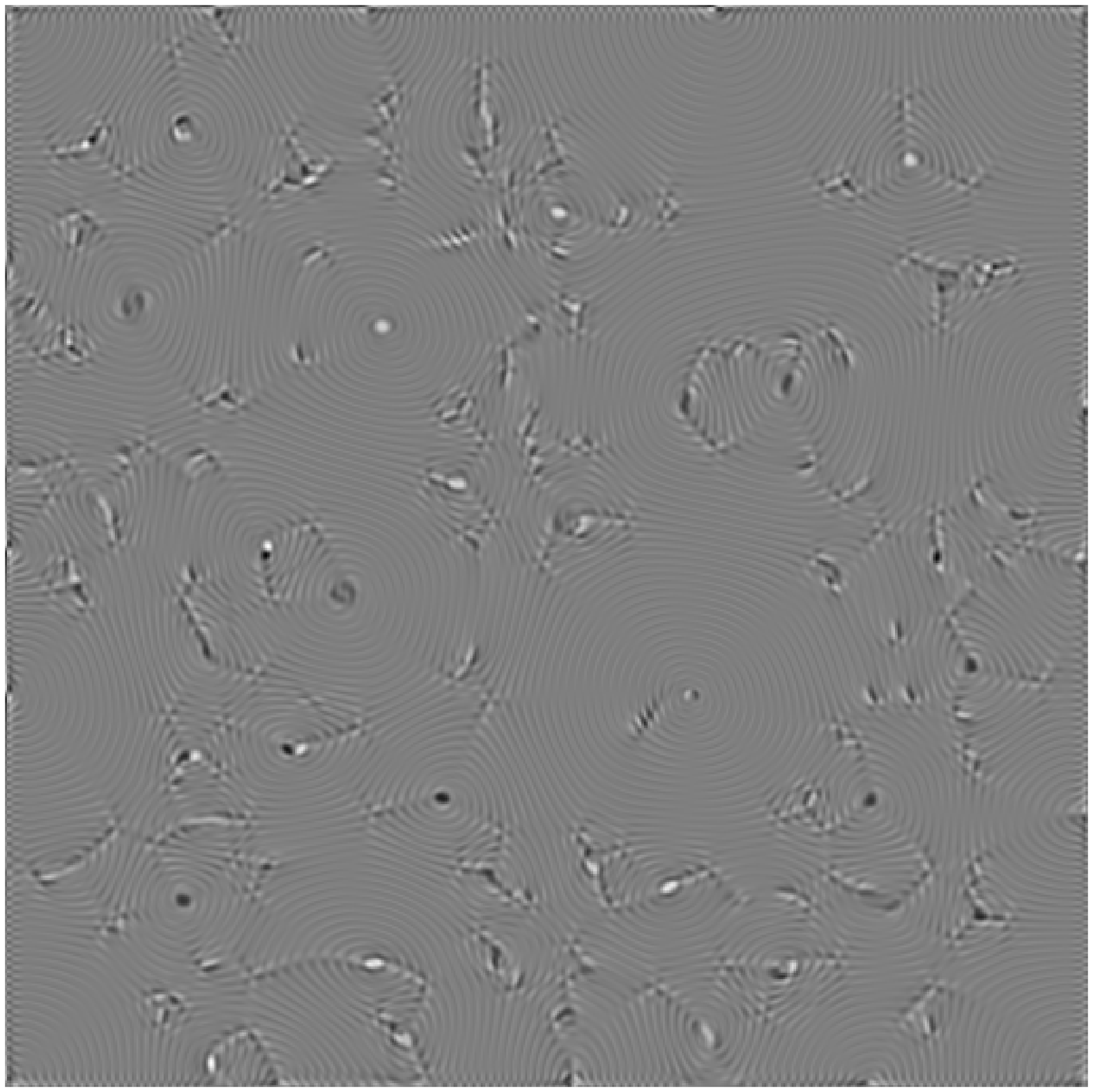} \\
(c) & (d) \\
\end{tabular}
\caption{Instantaneous patterns of $\psi(\rr,t)$
and $\omega_z(\rr,t)$. Dark regions correspond to $\psi(\rr,t) > 0$ 
or $\omega_z(\rr,t) > 0$ while  white regions to $\psi(\rr,t) < 0$ or
$\omega_z(\rr,t) < 0$.  
(a) $\psi(\rr,t)$ and (b) $\omega(\rr,t)$ at $\epsilon = 0.2$ on 
the roll branch; 
(c) $\psi(\rr,t)$ and (d) $\omega(\rr,t)$ at $\epsilon = 0.65$ on 
the SDC branch.} 
\lbl{fig_img1}
\end{figure}

	The patterns we observed are very similar to those found in real
experiments \cite{mo_bo_93,as_st_93}, which can be summarized 
as the following: 
(A) Within the roll branch: Straight parallel rolls are observed
at $\epsilon = 0.05$ up to $0.4$ with a few defects at the boundary.
Starting at $\epsilon = 0.45$  up to $0.6$,
the rolls start to bend and focal singularities start to
appear near the boundary. Weak time-dependence sets in at $\epsilon = 0.45$
in which the nucleation of defects along the sidewall occurs.
Two typical shadow graph images,
one for the order parameter $\psi(\rr,t)$ and another for the vertical
vorticity $\omega_z(\rr,t) = - \lpl \zeta(\rr,t)$,
of the roll state at $\epsilon = 0.2$ are shown in Fig. 2.
(B) Within the SDC branch: SDC states are observed at $\epsilon = 0.8$
to $0.55$, whose behavior has
been described in detail before \cite{mo_bo_93,as_st_93,xi_gu_95}.
At $\epsilon = 0.5$, only a few spirals still exist which 
mix with a background of locally curved rolls.
Finally at $\epsilon = 0.4$, the pattern looks much
like a roll state with a few defects and dislocations.
In Fig. 2, we plot two typical
shadow graph images, again one for $\psi(\rr,t)$ and another for 
$\omega_z(\rr,t)$,
of SDC at $\epsilon = 0.65$. While the vertical vorticity
$\omega_z(\rr)$ in a roll state is almost
zero everywhere, the corresponding field has a much richer structure in SDC.
This suggests choosing $\omega_z(\rr,t)$ to be an order parameter
in distinguishing a roll state from a SDC state \cite{xi_gu_95}. 
It is interesting to point out that patterns in the interval 
$\epsilon = 0.4 - 0.6$
depend on their earlier histories, i.e., whether they are on the roll branch
or the SDC branch; see Fig. 3. 
So a hysteretic loop exists when one follows the two
different routes. This is consistent with experimental observations that 
different patterns evolve
from different initial conditions at the same $\epsilon$ 
\cite{ca_eg_97}. 
Owing to the existence of hysteresis, 
the transition temperature
$\epsilon_T$ between  parallel roll states and SDC states
cannot be determined precisely in our study. Our rough
estimate is $\epsilon_T \simeq 0.45$.

\begin{figure}[t]
\centering
\begin{tabular}{cc}
\epsfxsize = \figsize \epsfysize = \epsfxsize \epsfbox{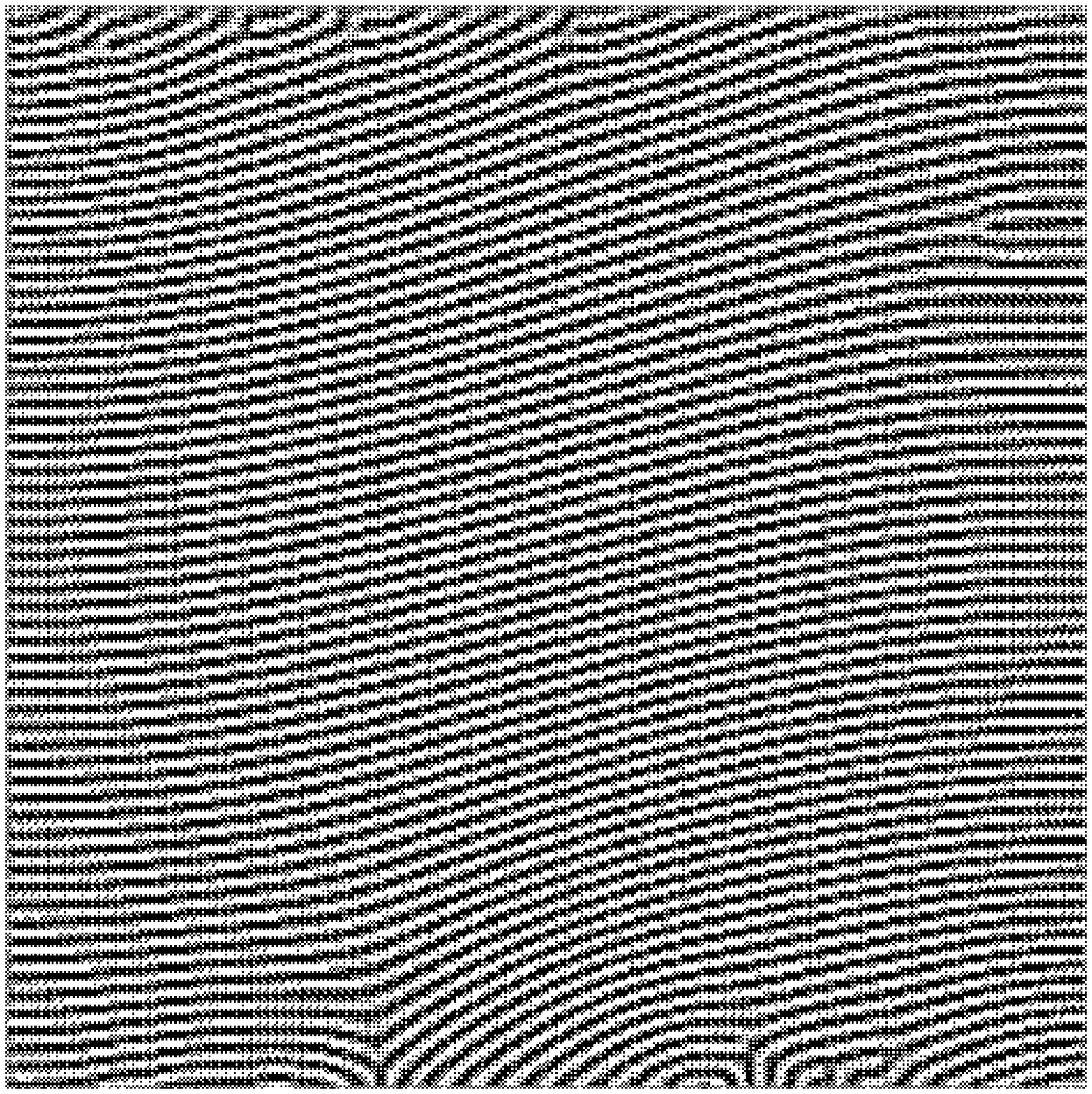} &
	\epsfxsize = \figsize \epsfysize = \epsfxsize 
	\epsfbox{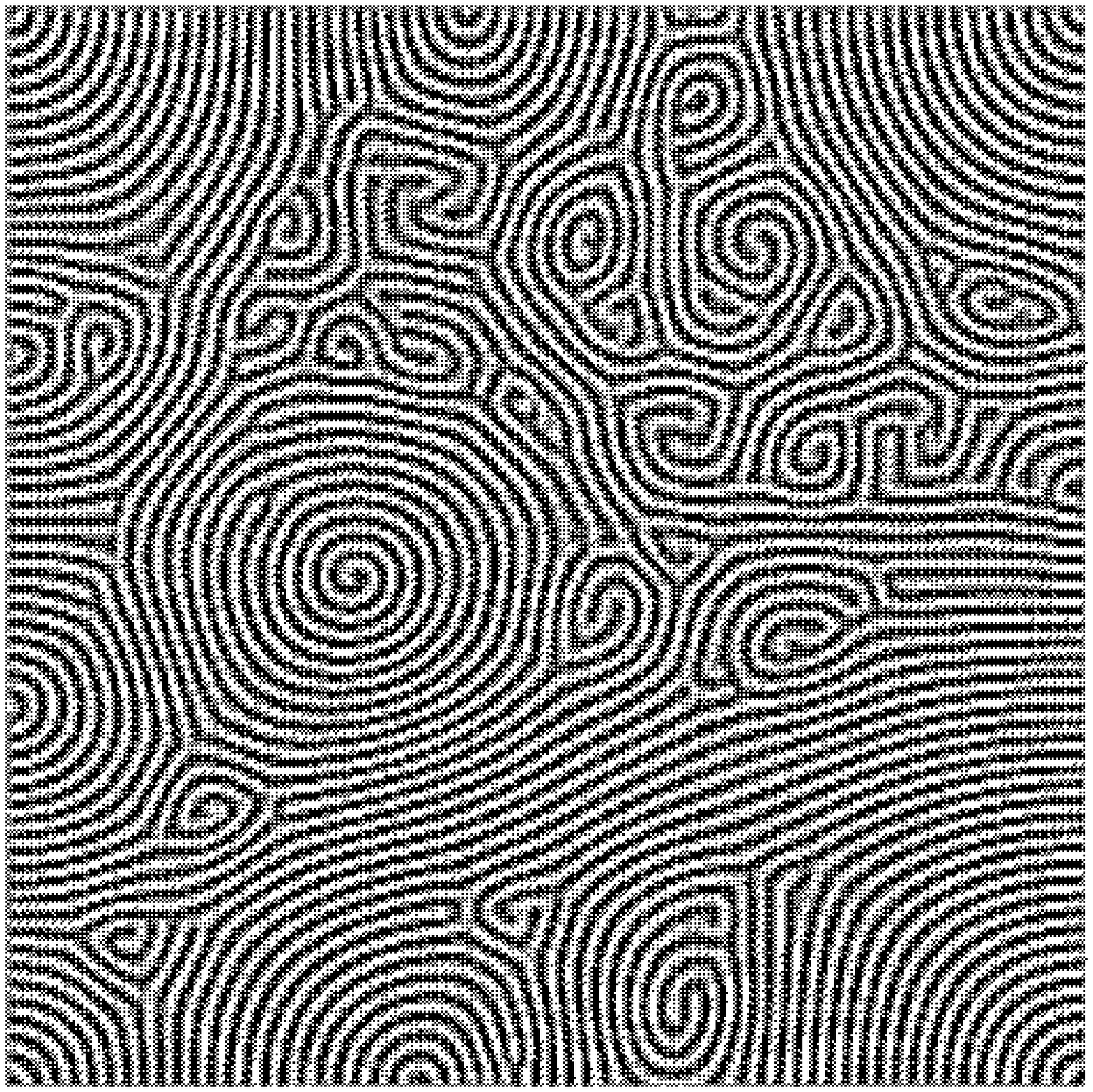} \\
(a) & (b) \\
\end{tabular}
\caption{Instantaneous patterns of $\psi(\rr,t)$
at $\epsilon = 0.55$: (a) on the roll branch and (b) on the SDC branch.
Dark regions correspond to $\psi(\rr,t) > 0$ 
while  white regions to $\psi(\rr,t) < 0$.}
\lbl{fig_img2}
\end{figure}

	In order to 
determine the character of the transition between rolls and SDC, we
have calculated, from our numerical studies,  
the structure factor $S(k)$, the two-point correlation length $\xit$, 
the convective current $J$, the vorticity current $\Omega$, 
and the spectra entropy $\Xi$
for both roll states and
SDC states. The numerical methods used in calculating
these quantities are the same as in Ref. \cite{xi_gu_95}.
The numerical uncertainties are taken as the variances of our data. 
Considering that we can, at most, take just a few samples of the 
strongly fluctuating instantaneous quantities, presumingly obeying 
Gaussian distributions near their corresponding time-averaged values, 
we believe that the probabilities, and hence the uncertainties, 
for us to obtain the truly time-averaged
values are determined by those variances. 
From the data for $S(k)$ of the SDC states,
we have verified  
the existence of the scaling form \rf{s_scaling}
within our numerical uncertainties for SDC \cite{li_xi_97}. 
The results for $\xit$, $J$, $\Omega$ and $\Xi$ are plotted in Figs. 4 - 7.
The most striking feature in these figures is the hysteretic loops
formed by the two branches.
It is also noticeable that the uncertainties
on the SDC branch are generally larger than those on the roll branch, which,
presumably, is due to the chaotic character of SDC.

	We fit the data on the roll branch with power-law behaviors. To
allow for the possibility that roll states might be unstable or metastable
for $\epsilon > 0.45$, only data for $0.05 \le \epsilon \le 0.45$ are used
in actual fittings. (i) We first use the nonlinear $\chi^2$ method to fit
the convective current with $J = J_0 (\epsilon - \epsilon_c)^\mu$ and
find that $J_0 = 0.6554 \pm 0.0002$, $\mu = 1.0054 \pm 0.0004$ and 
$\epsilon_c = 0.002$. The actual accuracy in our results 
may not as good as indicated. 
The fitting error for $\epsilon_c$ is very small.  This 
$\epsilon_c$ is the measured onset from conduction to convection, whose 
positive value 
is most likely due to finite-size effects \cite{wa_ah_81}. 
Apparently $J$ on the roll branch has a mean-field exponent. The
amplitude $J_0$ is also in good agreement with Eq. \rf{J_2} provided
that $\pv{x^2}_x \ge 1$ is not too big. 
(ii) Using the $\chi^2$ method, we fit the data for the correlation length 
with $\xit = \xi_{2,0} (\epsilon - \epsilon_c)^{-\nu}$, which leads to
$\xi_{2,0} = 13.0 \pm 0.2$ and $\nu = 0.54 \pm 0.01$.
So $\xit$ on the roll branch also has a mean-field exponent. 
(iii) Using the $\chi^2$ method, we fit the data for the vorticity current
with $\Omega = \Omega_0 (\epsilon - \epsilon_c)^\lambda$ and find that 
$\Omega_0 = (1.952 \pm 0.008) \times 10^{-9}$ and $\lambda = 2.461 \pm 0.003$. 
Again, the actual accuracy in our results 
may not as good as indicated. 
This behavior of $\Omega$ is not easy to understand.  
The amplitude equations coupled with mean-flow 
predict, for rigid-rigid boundaries, 
that $\Omega \sim \epsilon^{7/2}$ for almost perfect 
parallel rolls and $\Omega \sim \epsilon^3$ for general patterns
\cite{si_zi_81}.
None of these can explain the behavior of $\Omega$ we found, which seems to
be consistent with $\Omega \sim \epsilon^{5/2}$.   
(iv) The analysis of the spectra entropy is most difficult since there is
no theory whatsoever to describe its behavior. We simply fit it to a form 
$\Xi = \Xi_b + \Xi_0 (\epsilon - \epsilon_c)^\delta$ 
for the roll states. We find from the nonlinear $\chi^2$ method 
that $\Xi_b = 4.24 \pm 0.04$, $\Xi_0 = 2.5 \pm 0.1$ and 
$\delta = 1.00 \pm 0.08$. The background term $\Xi_b$ is much larger than
the corresponding value for perfect parallel rolls
$\Xi_b = \ln 2 = 0.6931$. 
This, presumably, is due to finite-size effects and/or limited 
computing time. 
The original data and their corresponding fitting curves for these global
quantities are plotted in Figs. 4 - 7.

The analyses of the data on the SDC branch 
must be treated with caution given that 
there are three possible scaling scenarios as discussed in 
Sec. \srf{sce}. 
We first fit the data of $\xit$ and $\Omega$ to power laws such as 
(i) $\xit = \xi_{2,0} (\epsilon - \epsilon_0)^{-\nu}$ 
and (ii) $\Omega = \Omega_0 (\epsilon - \epsilon_0)^\lambda$, 
where $\epsilon_0 = \epsilon_c$ [the onset temperature in a finite system] 
in scenario (A) or $\epsilon_0 = \et$ [the roll-to-SDC transition temperature] 
in scenario (B). 
Then we fit the data of $J$ 
in accordance with 
our theoretical result in Eq. \rf{tvcvcrnt}, i.e., 
(iii) $J = J_0 (\epsilon - \epsilon_c) - J_\xi \xit^{-2}$, where 
we take the corresponding fitting results for $\xit$ in (i). 
Clearly the value of $\epsilon_0$ is essential to determine 
which of the three scenarios is valid. 
In the following, we apply three different fittings for 
three values of $\epsilon_0$ and 
check the consistence of our numerical data against 
the theoretical
results in Eqs. \rf{tvcvcrnt} and \rf{tvvtctcrnt}.

(a) To check whether
scenario (A), i.e., $\te = \epsilon$, is valid, 
we fix $\epsilon_0 = \epsilon_c = 0.002$, whose value is given by
the fitting of $J$ on the roll branch. Since
it is probable that SDC states are unstable
or metastable at $\epsilon = 0.4$ and $0.45$, the corresponding data 
might deviate from their ``real'' values in order to form the
hysteretic loops. For this reason, we disregard these two points and
use only those data within 
$0.5 \le \epsilon \le 0.8$ for fitting. 
We use the $\chi^2$ method to fit our data, which leads to
(i) $\xi_0 = 6.8 \pm 0.2$ and $\nu = 0.72 \pm 0.05$; 
(ii) $\Omega_0 = (3.0 \pm 0.2) \times 10^{-8}$ and $\lambda = 3.0 \pm 0.1$; 
and (iii) $J_0 = 0.64 \pm 0.02$ and $J_\xi = 2.9 \pm 0.9$. 
The original data of $\xit$, $J$ and $\Omega$          
and their corresponding fitting curves are plotted in Fig. 4.
Apparently those curves fit the original data well. 
So scenario (A) is consistent with our numerical data. 
To further distinguish scenario (A1) [$\nu = \hf$] 
or (A2) [$\nu > \hf$], we find that, on the one hand, $\nu = 0.72 \pm 0.05$ 
from the direct fitting in (i) 
but, on the other hand, $\nu = 0.50 \pm 0.05$
from $\lambda = 3.0 \pm 0.1$ in (ii) and
$\lambda = 2 + 2 \nu$ in Eq. \rf{a_expnts}.
This discrepancy is likely caused by the 
big numerical uncertainties in our data. 
We feel that the direct fitting
is more reliable and scenario (A2)
is more likely to be true.
But we cannot definitely rule out scenario (A1).
More accurate data are needed to resolve this issue.  

(b) We check whether scenario (B), i.e., $\te = \epsilon - \et$,
is consistent with our numerical data, where the value of $\et$ 
is determined by the fitting of $\xit$. In contrast to case (a), 
there is no obvious reason
to disregard any point in this scenario. So   
all the data within $0.4 \le \epsilon \le 0.8$ are 
used in our fittings. 
We first use the nonlinear $\chi^2$ method to fit the data of $\xit$,  
which gives (i) $\xi_0 = 5.9 \pm 0.2$, 
$\nu = 0.46 \pm 0.06$ and $\epsilon_0 = \epsilon_T = 0.27 \pm 0.03$. 
Then, we fix 
$\epsilon_0 = 0.27$ and use the $\chi^2$ method to fit the data of 
$\Omega$ and $J$. We find that 
(ii) $\Omega_0 = (8.1 \pm 0.2) \times 10^{-8}$ 
and $\lambda = 2.33 \pm 0.01$; and 
(iii) $J_0 = 0.675 \pm 0.001$ and $J_\xi = 4.9 \pm 0.1$.  
These results are very sensitive to the points at 
$\epsilon = 0.4$ and $0.45$. The original data of $\xit$, $J$ and $\Omega$
and their corresponding fitting curves are plotted in Fig. 5.
The fitting of $\Omega$ obviously is not good. 
The results $\nu = 0.46 \pm 0.06$
in (i) and $\lambda = 2.33 \pm 0.01$ in (ii) do not satisfy 
the scaling relation $\lambda = 2 \nu$ in Eq. \rf{b_expnts}.  
So scenario (B) with $\et = 0.27$ is unlikely to be true. 

(c)  We check whether scenario (B), with $\et$ 
determined by the fitting of $\Omega$, is consistent with 
our numerical data.
As in case (b), we use all the data within $0.4 \le \epsilon \le 0.8$ in our
fittings. 
We first use the nonlinear $\chi^2$ method to fit the data of $\Omega$
and find that (ii) $\Omega_0 = (4.5 \pm 0.3) \times 10^{-8}$, 
$\lambda = 1.41 \pm 0.07$ and $\epsilon_T = 0.348 \pm 0.005$. 
Then we fix $\epsilon_0 = 0.348$ and 
apply the $\chi^2$ method to fit the data of $\xit$ and $J$. 
We find that (i) $\xi_{2,0} = 6.55 \pm 0.09$ and $\nu = 0.295 \pm 0.008$;
and (iii) $J_0 = 0.667 \pm 0.001$ and $J_\xi = 4.6 \pm 0.1$.
Again, these fitting results are very sensitive to the points at
$\epsilon = 0.4$ and $0.45$. The original data of $\xit$, $J$ and $\Omega$
and their corresponding fitting curves are plotted in Fig. 6.
Notice that, since $\nu$ in (i) is less than $\hf$, the slope of $J$ is 
negative in the range of $0 < \te = \epsilon - \et < \te_\times$ with
$\te_\times = [2 \nu J_\xi/\xi_{2,0}^2 J_0]^{1/(1-2 \nu)}$.
In the present case, one has $\te_\times = 3.20 \times 10^{-3}$, which perhaps 
is too small to be checked by real experiments or simulations.
From the pure-data-fitting point of view, the fittings in this case 
are as good as the fittings in case (a). 
So, without the benefit of our theoretical results, scenario (B) with 
$\et = 0.348$ could be (incorrectly) accepted. But, the exponents
$\nu = 0.295 \pm 0.008$ in (i) and $\lambda = 1.41 \pm 0.07$ in (ii) are not 
even close 
to satisfying the scaling relation $\lambda = 2 \nu$ in Eq. \rf{b_expnts}.  
So this scenario can be ruled out by our theory. From this, together with the
discussions in case (a) and (b), we conclude that scenario (B)
is unlikely to be valid for SDC and 
scenario (A) is consistent 
with our numerical data and our theory. 

\begin{figure}[t]
\begin{tabular}{r}
\epsfxsize = 0.93\plotsize
\epsfysize = 0.45\plotsize
\hbox{\raisebox{0.9\epsfysize}{$\xit^{-2}$}} \hskip -0.1in 
\epsfbox{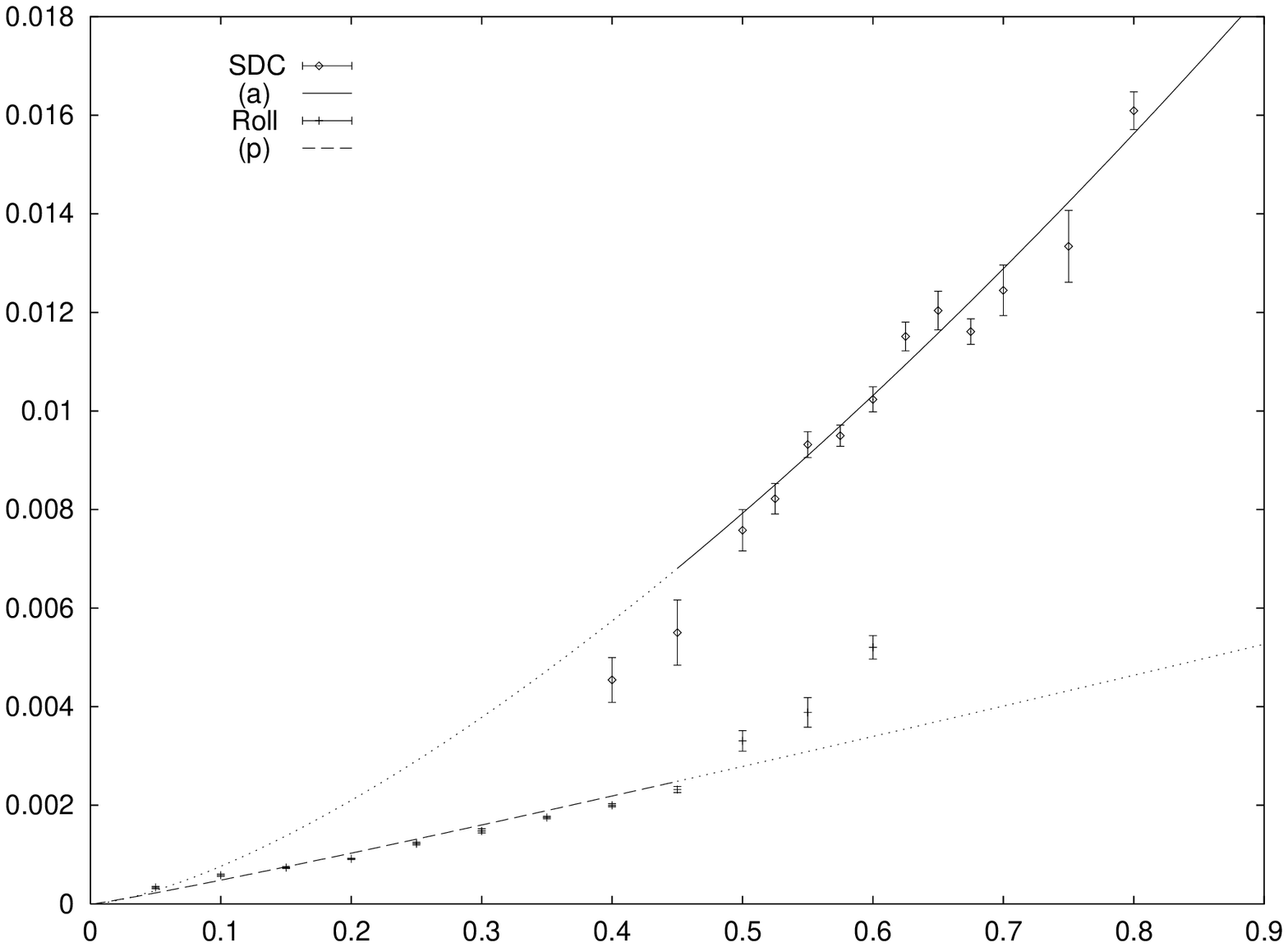}\\  
\epsfxsize = 0.93\plotsize
\epsfysize = 0.45\plotsize
\hbox{\raisebox{0.9\epsfysize}{$J$}} \hskip 0.in 
\epsfbox{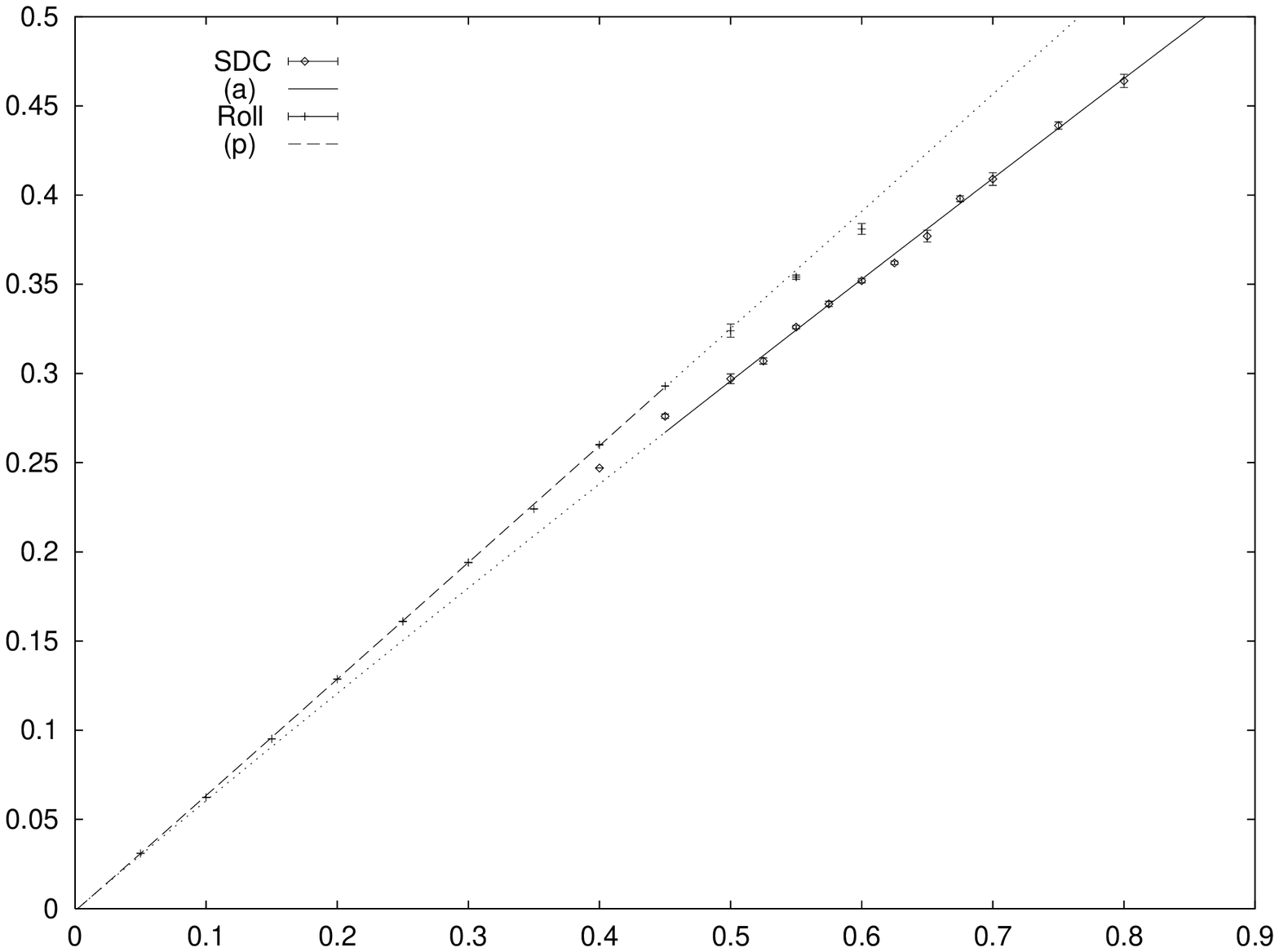}\\ 
\epsfxsize = 0.93\plotsize
\epsfysize = 0.45\plotsize
\hbox{\raisebox{0.9\epsfysize}{$10^9\Omega$}} \hskip -0.2in 
\epsfbox{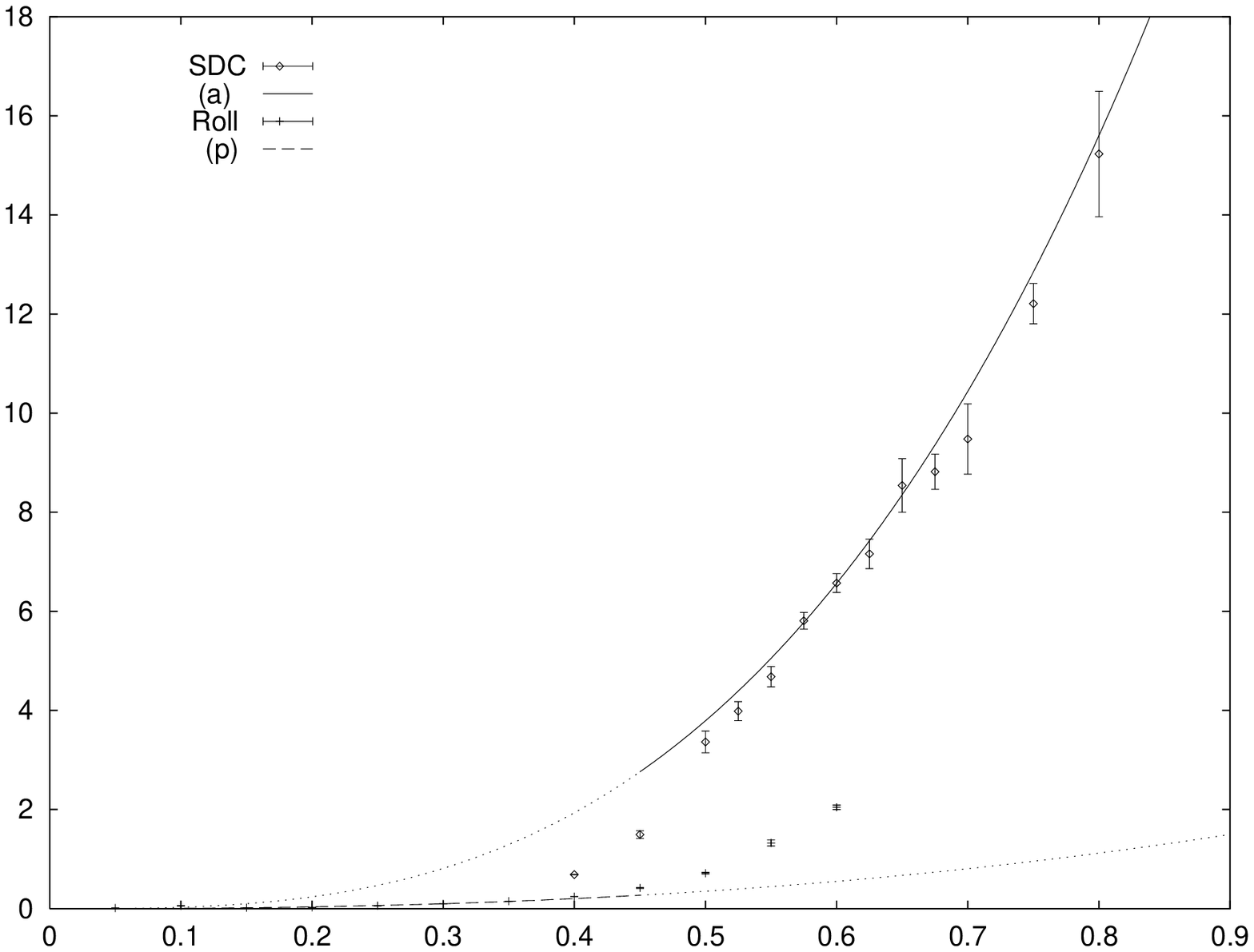}\\
\end{tabular}

\hskip 0.56\plotsize $\epsilon$
\caption{Plots of $\xit^{-2}$, $J$ and $\Omega$ 
vs. $\epsilon$ for both parallel rolls and SDC. The corresponding fitting
curves for SDC are described in (a) in Sec. IV. 
The labels are used  as the following: 
SDC --- the numerical data with error bars on the SDC branch; 
(a) --- the fitting curves for SDC described in (a) in the text;
Roll --- the numerical data with error bars on the roll branch; 
(p) --- the fitting curves for parallel rolls. 
To indicate the parallel roll states for $\epsilon > \et$ and
the SDC states for $\epsilon < \et$ may be metastable, the corresponding 
fitting curves in (a) and (p) are plotted with dotted lines. The
transition temperature is estimated roughly at $\et = 0.45$.}
\lbl{fig_a}
\end{figure}

It is worthwhile to mention that, 
in all cases above, the value of $J_0$
agrees with the theoretical result $2/3$ in Eq. \rf{tvcvcrnt};
the value of $J_\xi$ is also consistent with the theoretical prediction
$J_\xi \ge 8/3$. On the other hand, the
theoretical results in Eqs. \rf{a1_omega_amp} and 
\rf{b_omega_amp}  predict that 
$\Omega_0 = 7.4 \times 10^{-4}$ in case (a),  
$\Omega_0 = 7.2 \times 10^{-5}$ in case (b), and   
$\Omega_0 = 9.7 \times 10^{-5}$ in case (c). 
All these predictions are several orders
larger than the corresponding numerical results. 
The reason for such big discrepancies is not clear to us.

We have also studied the behavior of the spectra entropy.
Owing to the lack of any theory, 
we simply fit the data of $\Xi$ to a form 
$\Xi = \Xi_b + \Xi_0 (\epsilon - \epsilon_0)^\delta$ for the SDC branch.  
We apply two different fittings: (a) We
fix $\epsilon_0 = \epsilon_c$ and $\Xi_b = 4.24$ [from the fitting for 
the roll branch],  and use the $\chi^2$ method
to fit those data within $0.5 \le \epsilon \le 0.8$,  which leads to 
$\Xi_0 = 4.79 \pm 0.03$ and $\delta = 0.18 \pm 0.02$. (b) We fix 
$\Xi_b = 4.24$ and use the nonlinear $\chi^2$ method for all data within
$0.4 \le \epsilon \le 0.8$, which gives that 
$\Xi_0 = 5.13 \pm 0.05$, 
$\delta = 0.12 \pm 0.01$ and $\epsilon_0 = 0.37 \pm 0.02$. 
We have also
tried other alternative fittings 
such as using the nonlinear $\chi^2$ method for all data within
$0.4 \le \epsilon \le 0.8$ and fixing $\epsilon_0 = 0.27$
[from (i) in (b)] or
$\epsilon_0 = 0.348$ [from (ii) in (c)], but
none of them gives a reasonable fit. 
The fitting curves in (a) and (b) 
and the original data of $\Xi$ are plotted in Fig. 7.
At this stage, the behavior of
$\Xi$ is the most unclear one  among all the time-averaged
global quantities defined in Sec. \srf{th}.

\begin{figure}[t]
\begin{tabular}{r}
\epsfxsize = 0.93\plotsize
\epsfysize = 0.45\plotsize
\hbox{\raisebox{0.9\epsfysize}{$\xit^{-2}$}} \hskip -0.1in 
\epsfbox{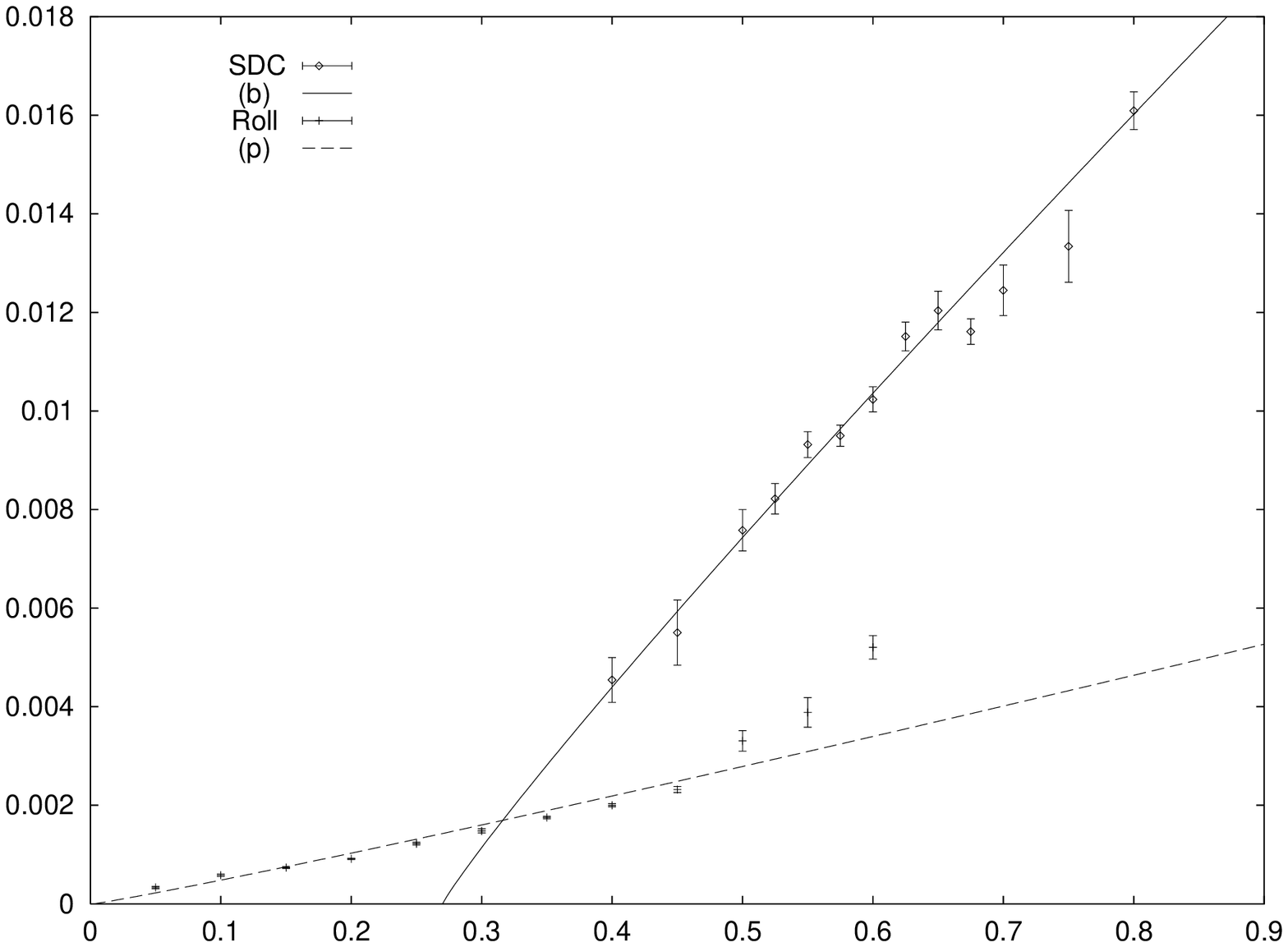}\\  
\epsfxsize = 0.93\plotsize
\epsfysize = 0.45\plotsize
\hbox{\raisebox{0.9\epsfysize}{$J$}} \hskip 0.in 
\epsfbox{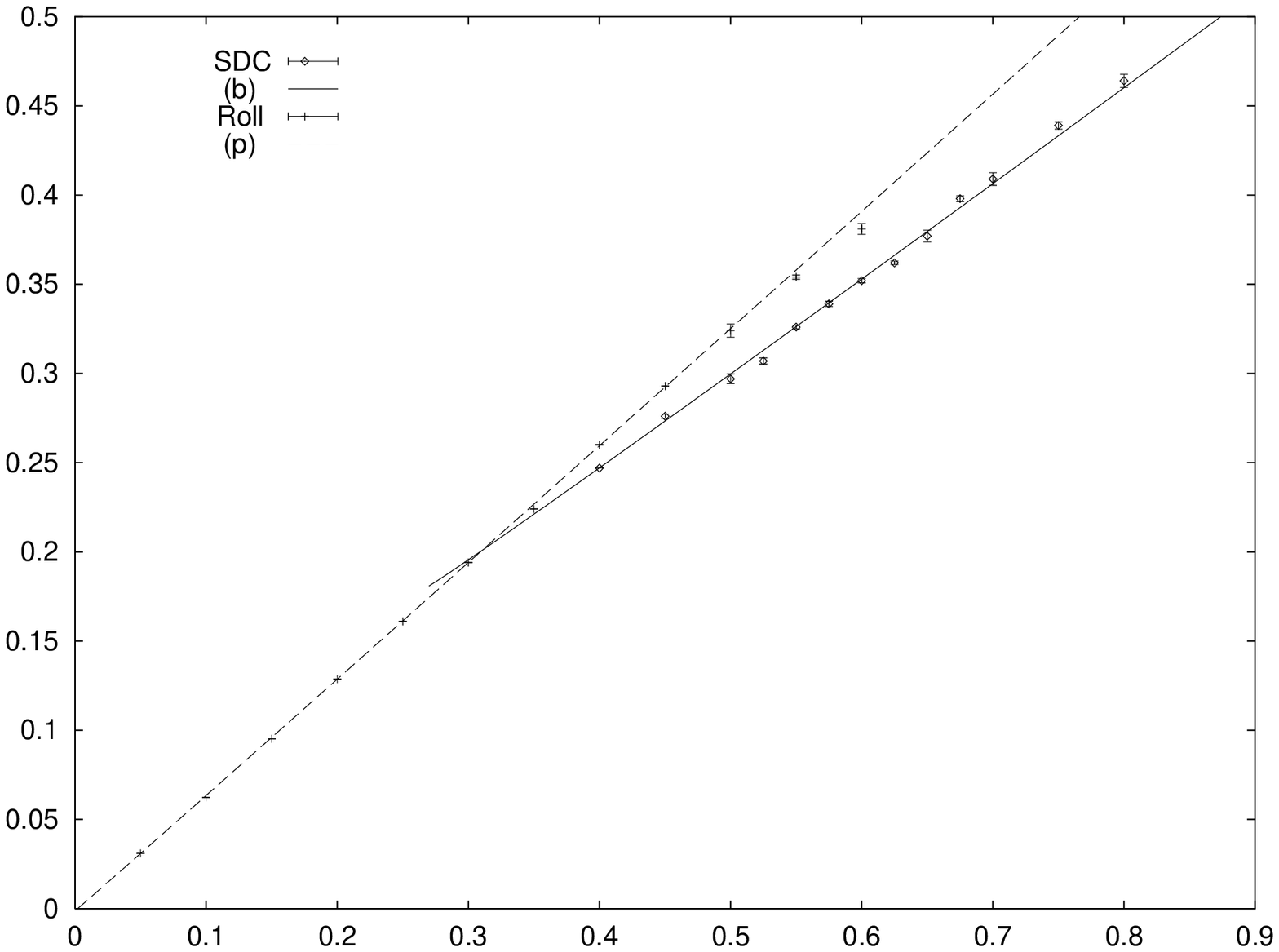}\\ 
\epsfxsize = 0.93\plotsize
\epsfysize = 0.45\plotsize
\hbox{\raisebox{0.9\epsfysize}{$10^9\Omega$}} \hskip -0.2in 
\epsfbox{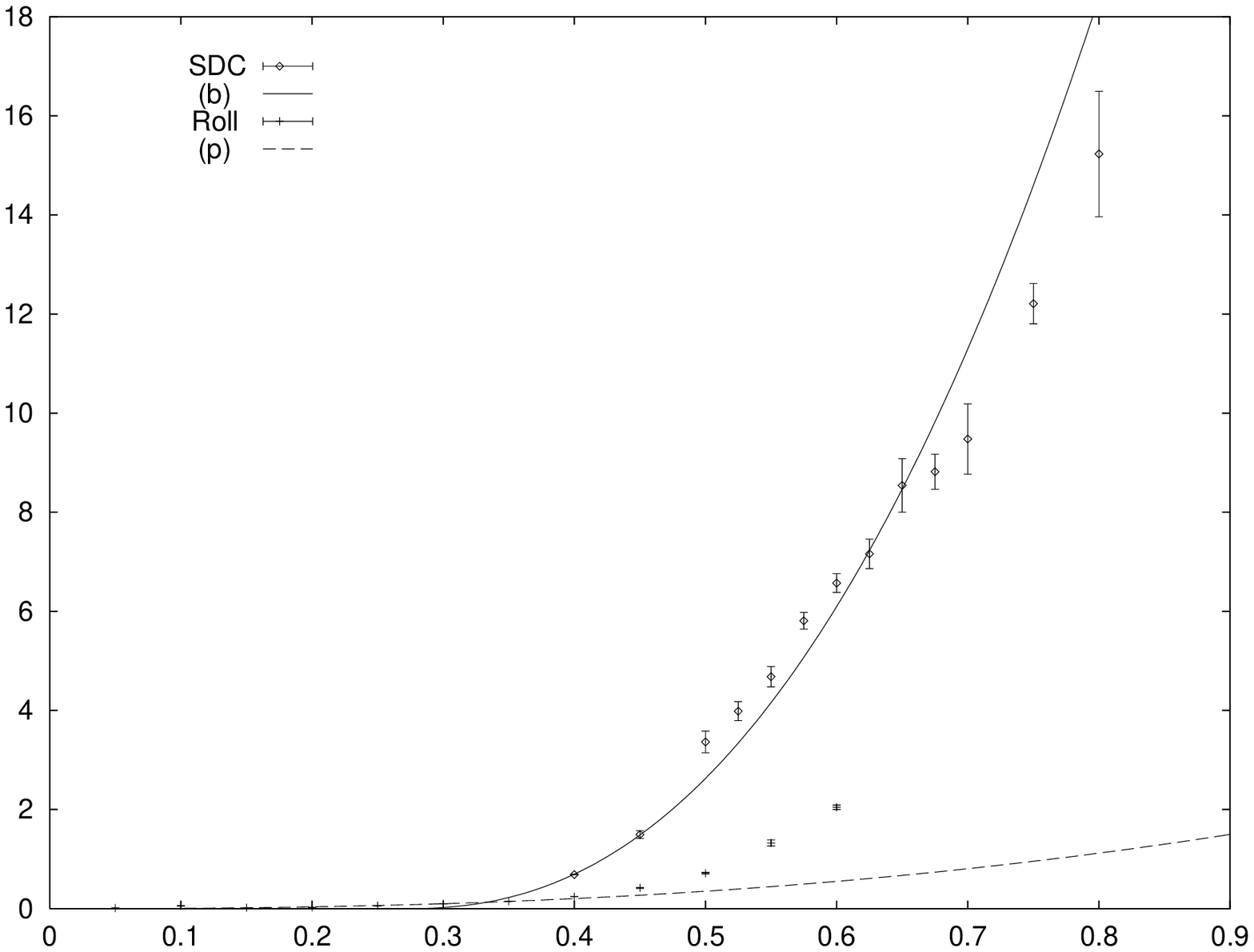}\\
\end{tabular}

\hskip 0.56\plotsize $\epsilon$
\caption{Plots of $\xit^{-2}$, $J$ and $\Omega$ 
vs. $\epsilon$ for both parallel rolls and SDC. The corresponding fitting
curves for SDC are described in (b) in Sec. IV. The labels are used  
as the following: 
SDC --- the numerical data with error bars on the SDC branch; 
(b) --- the fitting curves for SDC described in (b) in the text;
Roll --- the numerical data with error bars on the roll branch; 
(p) --- the fitting curves for parallel rolls.} 
\lbl{fig_b}
\end{figure}

\begin{figure}[t]
\begin{tabular}{r}
\epsfxsize = 0.93\plotsize
\epsfysize = 0.45\plotsize
\hbox{\raisebox{0.9\epsfysize}{$\xit^{-2}$}} \hskip -0.1in 
\epsfbox{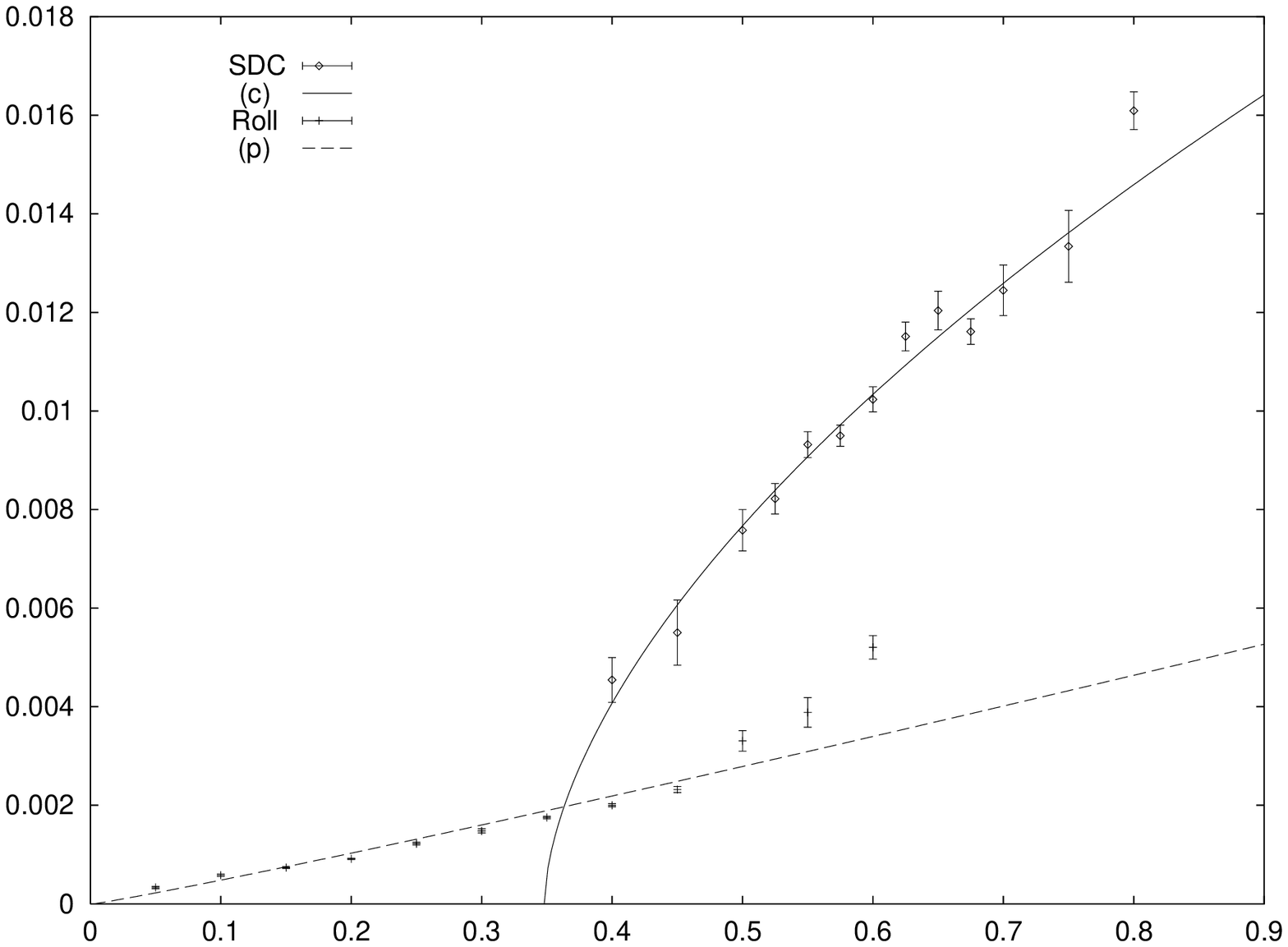}\\  
\epsfxsize = 0.93\plotsize
\epsfysize = 0.45\plotsize
\hbox{\raisebox{0.9\epsfysize}{$J$}} \hskip 0.in 
\epsfbox{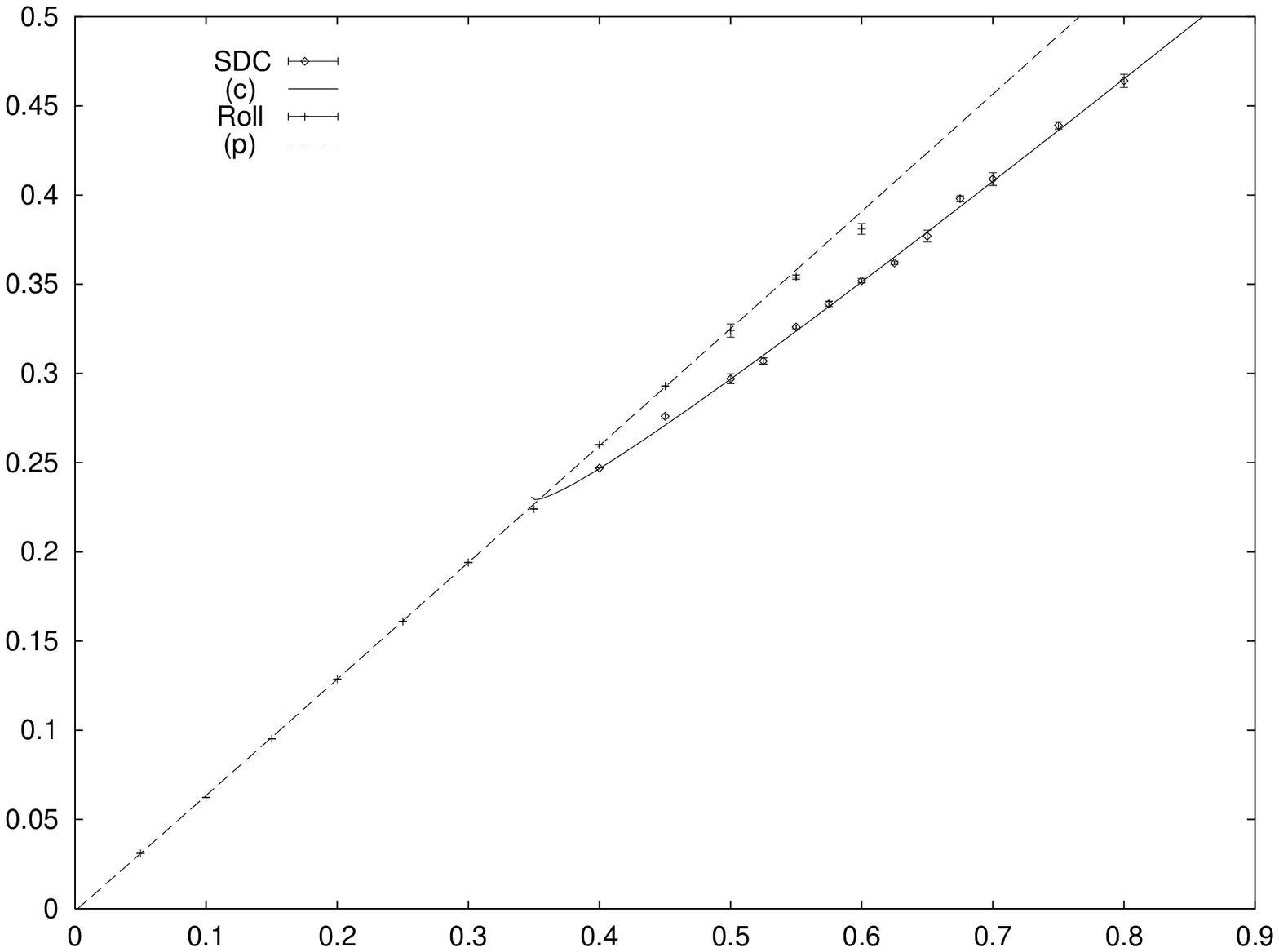}\\ 
\epsfxsize = 0.93\plotsize
\epsfysize = 0.45\plotsize
\hbox{\raisebox{0.9\epsfysize}{$10^9\Omega$}} \hskip -0.2in 
\epsfbox{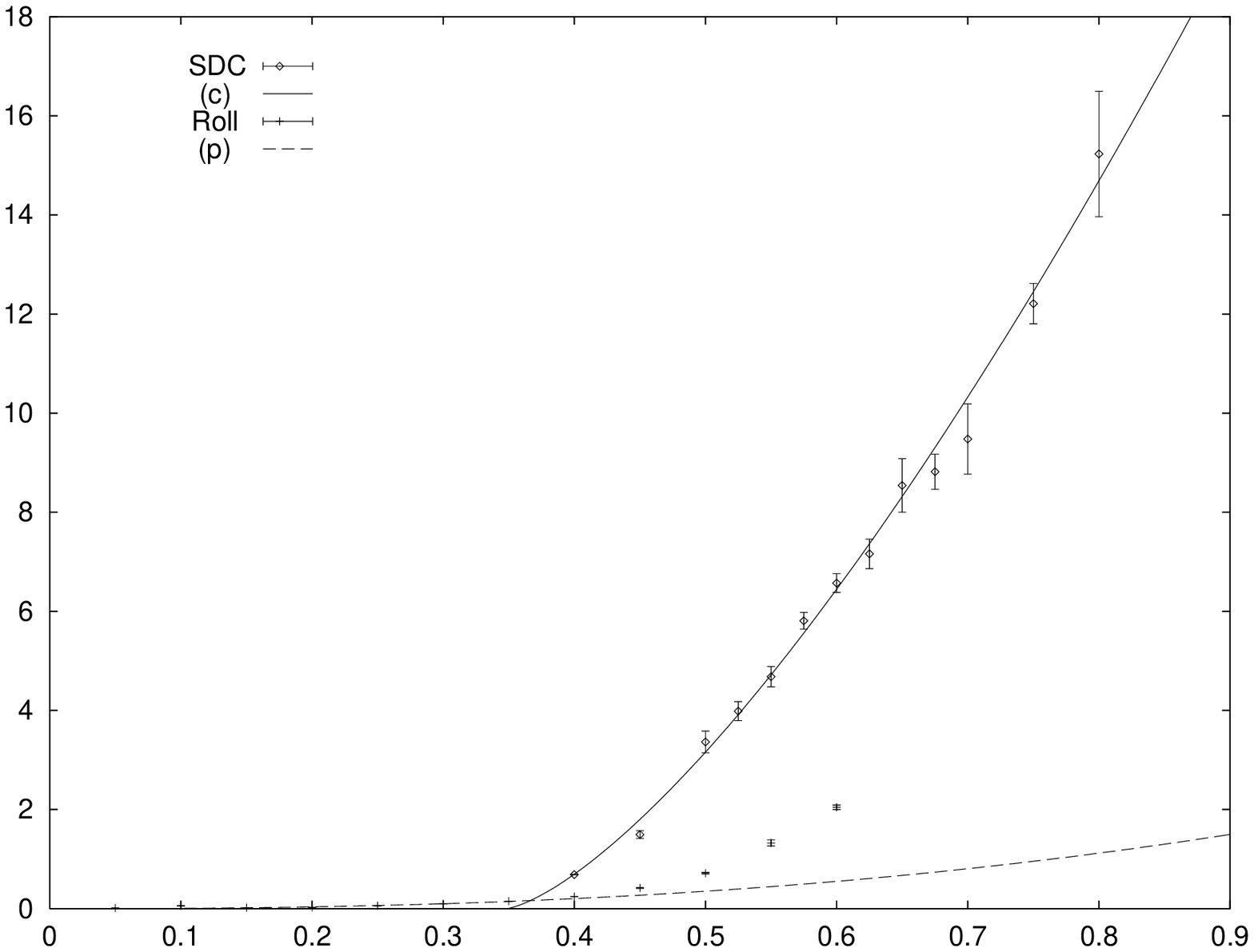}\\
\end{tabular}

\hskip 0.56\plotsize $\epsilon$
\caption{Plots of $\xit^{-2}$, $J$ and $\Omega$ 
vs. $\epsilon$ for both parallel rolls and SDC. The corresponding fitting
curves for SDC are described in (c) in Sec. IV. The labels are used  
as the following: 
SDC --- the numerical data with error bars on the SDC branch; 
(c) --- the fitting curves for SDC described in (c) in the text;
Roll --- the numerical data with error bars on the roll branch; 
(p) --- the fitting curves for parallel rolls.} 
\lbl{fig_c}
\end{figure}

\section{Discussion and Conclusion} \lbl{conc}

In the previous section, we conclude that scenario (A) 
is valid for SDC. This means that 
$\xit$ diverges at $\epsilon = 0$, 
and $J$ and $\Omega$ vanish at $\epsilon = 0$. At first sight 
it seems puzzling that all properties of SDC are controlled by $\epsilon$, 
instead of $\epsilon - \epsilon_T$. To understand this,
we propose an explanation for this scenario, which is somewhat similar
to that in the hexagon-to-roll transition in non-Boussinesq fluids 
\cite{bu_67}. In the latter case, the transition from hexagonal states
to parallel roll states occurs at finite $\epsilon_T$. 
Although the roll attractor is 
unstable for small enough $\epsilon$ and metastable against the
hexagonal attractor
for even slightly larger $\epsilon$, the properties of the parallel 
roll states are all 
controlled by $\epsilon$, not by $\epsilon - \epsilon_T$ \cite{bu_67}.
Clearly one can imagine a similar picture for the roll-to-SDC transition.
While the SDC attractor seems to be either unstable or 
metastable against the roll attractor for sufficiently 
small $\epsilon$, as an intrinsic convective state, the 
properties of SDC
are controlled at the conduction to convection threshold, not where
it starts to emerge as the stable state. 
The existence of two different attractors has  
been suggested by experiments \cite{mo_bo_96,ca_eg_97}. The basins and the
stability of these two attractors are still unclear at present. 

\begin{figure}[t]
\epsfxsize = \plotsize
\epsfysize = 0.67\plotsize
\centering
\hbox{\raisebox{0.93\epsfysize}{$\Xi$} \hskip -0.1in
\epsfbox{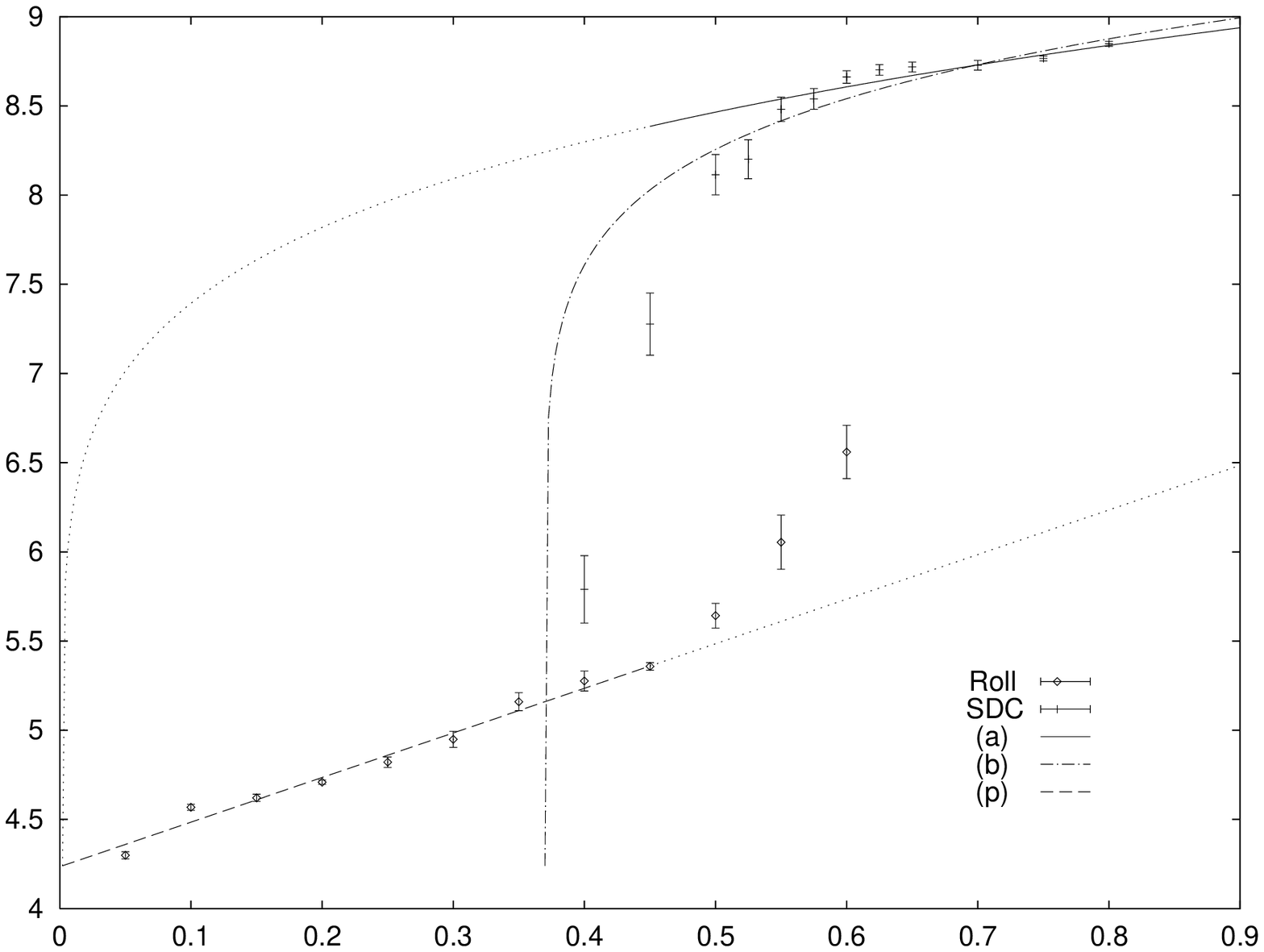}}

\hskip -0.05\plotsize
$\epsilon$
\caption{Plot of $\Xi$ vs. $\epsilon$ for both
parallel rolls and SDC. 
The different fitting forms (a) and (b) for SDC are discussed in the text.
The labels are used  as the following: 
Roll --- the numerical data with error bars on the roll branch; 
SDC --- the numerical data with error bars on the SDC branch; 
(a) --- the fitting curve for SDC described in (a) in the text;
(b) --- the fitting curve for SDC described in (b) in the text;
(p) --- the fitting curve for parallel rolls. 
To indicate the parallel roll states for $\epsilon > \et$ and
the SDC states for $\epsilon < \et$ may be metastable, the corresponding 
fitting curves in (a) and (p) are plotted with dotted lines. The
transition temperature is estimated roughly at $\et = 0.45$.}
\lbl{fig_entropy}
\end{figure}

The establishment of scenario (A) indicates that 
the transition between the 
parallel roll states and the SDC states is first-order. This conclusion 
is also supported by the following:

(1) Our theory predicts discontinuities in the value and the slope
of $J$ at $\epsilon_T$
[see the discussion following Eq. \rf{d_j}].  
This is a typical signature of a first-order transition. 

(2) The 
presence of hysteretic loops in Figs. 4 - 7 
is a strong indication of a first-order transition. 
A different hysteretic loop has also been reported
by others for the GSH model \cite{cr_tu_95}. 
Although it is arguable
that hysteretic loops may be found in a second-order transition if the 
computing time is not long enough to overcome the effects of 
critical slowing-down (which 
occurs when the correlation time $\tau$ approaches infinity), 
it is doubtful that the loops in that case
can be as distinctive as what we found in 
Figs. 4 - 7.

(3) As we described in Sec. \srf{num}, the convective patterns depend
on the processes leading to them, which is also observed in experiments 
\cite{ca_eg_97}.  This fact suggests that the two competing
attractors are either 
both stable or one is stable while the other is metastable for some positive
$\epsilon$. Such a stability property 
is typical in first-order transitions. On the contrary, 
in second-order transitions one of the two
attractors should change from stable to unstable while the other changes
from unstable to stable as $\epsilon$ moves 
across $\et$. 

(4) From Figs. 5 - 7, it is easy to see that, 
if scenario (B) is valid, then the fitting curves of all the time-averaged
global quantities on the SDC branch will cross those on the roll branch. 
But there is no evidence from our numerical calculation supporting such a
crossing.  

As we mentioned in Sec. \srf{int}, an earlier 
experiment with a circular cell \cite{mo_bo_96}
found that $\xit$ diverges at $\epsilon = 0$ with a mean-field exponent, while
the correlation time $\tau$
either diverges at $\epsilon = 0$ with a non-mean-field
exponent or diverges at the roll-to-SDC transition temperature $\et$ with a
mean-field exponent. 
However, a recent experiment with a square cell \cite{ca_eg_97} 
concluded that $\xit$ diverges at $\et$ with a very small exponent $\nu$. 
We now comment on these experiments and our study.

Regarding our numerical study, 
we cannot rule out
that the roll-to-SDC transition in the GSH model has a different character
from those in real experiments  although this seems unlikely. 
We also cannot rule out that our numerical
solutions are still in the transient regime even though we have waited for
about four horizontal diffusion times $t_h$
before collecting data over an interval of
several $t_h$ for each $\epsilon$. Furthermore, 
as we discussed in Sec. \srf{num}, 
our numerical data are not accurate enough 
to determine independently which of the three scenarios is true for SDC. 
As a result, 
we have to rely on our theoretical predictions to resolve this issue.
Finally we disregarded two data points in our analysis for scenario (A) on the
basis that these two points deviate from their ``real'' values
to form hysteretic loops. This 
introduces a certain arbitrariness in determining
which data points deviate.  
These shortcomings in our numerical
study weaken the validity of our conclusion.  

We notice that the data of the correlation length $\xit$
for parallel rolls and SDC were analyzed together in the experiment
by Morris {\it et al.} \cite{mo_bo_96}, 
which we think is not justified. 
Considering that the parallel roll states and the SDC states are
intrinsically different, we believe it is necessary to 
separate their data in the analysis, such as  we did in 
Sec. \srf{num}. Such a separation was implicit for the data of the
correlation time $\tau$ since $\tau = +\infty$ for steady states
such as parallel rolls and, in principle,  
only the data for SDC are available.  
A divergence at $\epsilon = 0$ with a non-mean-field exponent  
was found to be consistent with the data of $\tau$ for SDC \cite{mo_bo_96}. 
It is not clear to us whether a similar conclusion can be reached for $\xit$ 
if its data for SDC are analyzed separately. 

The conclusion by Cakmur {\it et al.} \cite{ca_eg_97} that
$\xit$ diverges at $\et$ with a small exponent $\nu$ is different from
that in Ref. \cite{mo_bo_96} and ours. These authors suggested that the
different geometry of experimental cells is the cause for the different
results between theirs and that in Ref. \cite{mo_bo_96}. 
If so, then it is not clear to us what the real behavior is in an infinite
system.  In our calculation,  we also used a square cell. 
We think that there may exist 
a different interpretation of the experimental data for $\xit$ obtained in 
Ref. \cite{ca_eg_97}.  
We believe that, to convincingly show that $\xit$ for SDC diverges at $\et$,
one must have sufficient data points whose $\xit$ 
are much {\it larger} than the corresponding typical 
values of parallel rolls (for $\epsilon$ away from both $0$ and $\et$). 
Otherwise, the data for $\xit$ for SDC may simply 
approach those for parallel rolls to form a hysteretic loop near $\et$, 
instead of diverging at $\et$. 
Since no data for $\xit$ for the parallel roll states was given
and the number of data for SDC near $\et$ was relatively limited,  
we feel the conclusion by Cakmur {\it et al.} \cite{ca_eg_97}
should be treated with caution. 

As we discussed in Sec. \srf{num}, our theory plays an important role in
determining the nature of the roll-to-SDC transition. So it is very
important to check our theoretical predictions by real experiments. 
One important prediction by our theory is that there exist 
discontinuities in the value and the slope of $J$ at $\et$.   
However, we realize that no 
discontinuity in $J$ has been reported by experiments. 
The reason for this is not clear to us.   
We conjecture that finite size effects might play a role. 
From Fig. 1, we find that the discontinuity is larger for smaller
Prandtl number $\sigma$. So it would be interesting to see whether experiments
can confirm or rule out such a discontinuity in $J$ by using a small $\sigma$. 
Another important prediction from our analysis is  
the behavior of the time-averaged vorticity current $\Omega$.  
Since direct measurements of $\Omega$ seem to be very 
difficult in real experiments \cite{cr_ga_86}, 
we think it valuable to {\it calculate} 
$\Omega$ by solving Eq. \rf{mf} [the corresponding version before rescalings
can be found in Ref. \cite{li_xi_97}] or its improved versions 
numerically, with the experimental results of $\psi(\rr,t)$ as input.   
Such a calculation will not only help to clarify the nature of the 
roll-to-SDC transition, 
but also provide an additional 
experimental test on our theory \cite{li_xi_97}. 
It would also be useful to 
calculate the time-averaged spectra entropy as suggested in
Refs. \cite{ca_eg_97,xi_gu_95}, even though 
there is no theory to predict the behavior of this quantity.

In summary, we conclude from our numerical studies and our
theoretical results that the roll-to-SDC transition is first-order
in character. We found that 
the correlation length $\xit$ for SDC diverges at $\epsilon = 0$,
not at the transition temperature $\et$. 
However, since the uncertainties in our data are unpleasantly
large and the data points we have are unsatisfactorily few in number, 
we cannot determine definitely whether the exponent of $\xit$ is mean-field
or not.  So further investigations are necessary to draw a definite
conclusion. In this regard, a theoretical calculation of $\xit$ for SDC
is highly desirable. A theory to describe the roll-to-SDC  transition is 
essential. Finite size effects should also be studied carefully.

\begin{center} Acknowledgment \end{center}

	X.J.L and J.D.G are supported by the National Science Foundation 
under Grant No. DMR-9596202. H.W.X. is supported by the Research 
Corporation under Grant No. CC4250. 
Numerical work reported
here was carried out on the Cray-C90 at the Pittsburgh
Supercomputing  Center and Cray-YMP8 at the Ohio Supercomputer Center.

\end{multicols} 

\begin{references}

\bm{cr_ho_93} 
	For a recent review on pattern formation in various systems, 
	see: M.~C.~Cross and
	P.~C.~Hohenberg, Rev. Mod. Phys. {\bf 65}, 851 (1993). 

\bm{gr_96}
	H. S. Greenside, chao-dyn/9612004.

\bm{ah_95} G.~Ahlers, in {\it 25 Years of Nonequilibrium
	Statistical Mechanics}, 
	edited by J. J. Brey {\it et al.} (Springer, New York, 1995), p. 91.

\bm{sc_lo_65}	
	A. Schl\"{u}ter, D. Lortz, and F. Busse, J. Fluid
		Mech. {\bf 23}, 129 (1965);
	F.~H.~Busse, Rep. Prog. Phys. {\bf 41}, 1929 (1978);
	F.~H.~Busse and R.~M.~Clever, in {\it New Trends in Nonlinear
	Dynamics and Pattern-Forming Phenomena}, edited by P.~Coullet and
	P.~Huerre (Plenum Press, New York, 1990), p. 37;
	and references therein.

\bm{mo_bo_93} 
	S. W. Morris, E. Bodenschatz, D. S. Cannell and G. Ahlers,
	Phys. Rev. Lett. {\bf 71}, 2026 (1993).

\bm{as_st_93}
	M. Assenheimer and V. Steinberg, Phys. Rev. Lett. {\bf 70}, 3888 
	(1993);  Nature {\bf 367}, 345 (1994).

\bm{xi_gu_93} 
	H.-W.~Xi, J.~D.~Gunton and J.~Vi\~{n}als, Phys. Rev. Lett. {\bf 71},
	2030 (1993). 

\bm{be_fa_93} M.~Bestehorn, M. Fantz, R. Friedrich, and H. Haken,
	Phys. Lett. A {\bf 174}, 48 (1993).

\bm{de_pe_94} W.~Decker, W.~Pesch and A.~Weber, Phys. Rev. Lett. {\bf 73}, 648
	(1994).


\bm{hu_ec_95}
	Y.~Hu, R.~E.~Ecke and G.~Ahlers, Phys. Rev. Letts. 
	{\bf 74}, 391 (1995);
	J. Liu and G. Ahlers, \ibid {\bf 77}, 3126 (1996).

\bm{mo_bo_96} 
	S.~W.~Morris, E. Bodenschatz, D. S. Cannell, and G. Ahlers,
	Physica D {\bf 97}, 164 (1996). 

\bm{ca_eg_97}
	R. V. Cakmur, D. A. Egolf, B. B. Plapp, and E. Bodenschatz,
	patt-sol/9702003.

\bm{cr_tu_95} M.~C.~Cross and Y.~Tu, Phys. Rev. Lett. {\bf 75}, 834 (1995).

\bm{li_xi_97}  
	X.-J. Li, H.-W. Xi, and J. D. Gunton, patt-sol/9706007. 

\bm{xi_gu_95}
	H.-W.~Xi and J.~D.~Gunton, Phys. Rev. E {\bf 52}, 4963
	(1995). A factor of $10^{-9}$ was missed for the quantity $\Omega$
	there. 

\bm{sw_ho_77} 
	J.~Swift and P.~C.~Hohenberg, Phys. Rev. A {\bf 15}, 319 (1977).

\bm{cr_80}
	M. C. Cross, Phys. Fluids {\bf 23}, 1727 (1980); 
	G. Ahlers, M. C. Cross, P. C. Hohenberg, and S. Safran, 
	J. Fluid Mech. {\bf 110}, 297 (1981). 

\bm{si_zi_81}
	E.~D. Siggia and A.~Zippelius, Phys. Rev. Lett. {\bf 47}, 835 (1981);
	M. C. Cross, Phys. Rev. A {\bf 27}, 490 (1983);
	P.~Manneville, J. Phys. (Paris) {\bf 44}, 759 (1983).

\bm{po_pe_79}
	G. C. Powell and I. C. Percival, J. Phys. A {\bf 12}, 2053 (1979).

\bm{bu_67}
	F. H. Busse, J. Fluid Mech. {\bf 30}, 625 (1967); 
	C. P\'erez-Garc\'{i}a,  E. Pampaloni and S. Ciliberto,
	in {\it Quantitative Measures of Complex Dynamical Systems}, 
	edited by N. B. Abraham and A. Albano
	(Plenum, New York, 1990), p. 405;
	E. Pampaloni, C. P\'erez-Garc\'{i}a, L. Albavetti and S. Ciliberto, 
	J. Fluid Mech. {\bf 234}, 393 (1992). 

\bm{xi_vi_92}
	H.-W.~Xi, J.~Vi\~{n}als and J.~D.~Gunton, Phys. Rev. A {\bf 46},
	R4483 (1992); H.-W.~Xi, J.~D.~Gunton and J.~Vi\~{n}als,
	Phys. Rev. E {\bf 47}, R2987 (1993); 
	X.-J. Li, H.-W. Xi, and J. D. Gunton, \ibid {\bf 54},
	R3105 (1996).

\bm{gr_cr_85}
	H. S. Greenside and M. C. Cross, Phys. Rev. A {\bf 31}, 2492 (1985).

\bm{bu_89}
	F. H. Busse, in {\it Advances in Turbulence 2}, Edited by 
	H.-H. Fernholz and H. E. Fiedler (Springer-Verlag, Berlin, 1989);
	F. H. Busse, M. Kropp, and M. Zaks, Physica D {\bf 61}, 94 (1992). 

\bm{xi_li_97} H.-W. Xi, X.-J. Li, and J. D. Gunton, Phys. Rev. Lett. {\bf 78},
	1046 (1997); and [to be submitted].


\bm{gr_co_84} 
	P. E. Bj{\o}rstad {\it et al.}, in {\it Elliptic Problem Solvers II},
	Edited by G. Birkhoff and A. Schoenstadt (Academic, Orlando, 1984), 
	p. 531; H.S. Greenside and W.M. Coughran Jr., 
	Phys. Rev. A {\bf 30}, 398 (1984).

\bm{wa_ah_81}
	R. W. Walden and G. Ahlers, J. Fluid Mech. {\bf 109}, 89 (1981);
	R. P. Behringer and G. Ahlers, \ibid {\bf 125},
	219 (1982).

\bm{cr_ga_86}
	V. Croquette, P. Le Gal, and Pocheau, Europhys. Lett. {\bf 1}, 
	393 (1986); F. Daviaud and A. Pocheau, \ibid {\bf 9}, 675 (1989).

\end{references}
\end{document}